\documentclass[
	fontsize=11pt,
	parskip=half,
]{scrartcl}

\usepackage[dvipsnames]{xcolor}
			\definecolor{dark-gray}{gray}{0.1}
			\definecolor{dark-blue}{RGB}{0,0,80}
			\definecolor{light-gray}{gray}{0.9}

\usepackage{tikz-cd}
	\usetikzlibrary{babel}
	\usetikzlibrary{positioning}

\usepackage[english]{csquotes}	
	\MakeOuterQuote{"}

\usepackage{graphicx}
\usepackage[format=plain, indention=3ex, labelsep=quad, labelfont=bf]{caption}

\usepackage{amsmath}
\usepackage{mathtools}	
	\mathtoolsset{showonlyrefs,mathic}
\usepackage{amssymb}	
\usepackage{aliascnt}	
\usepackage{amsthm}	
	\theoremstyle{definition}
		\newtheorem{definition}{Definition}
		
		\newtheorem*{definition*}{Definition}
	\theoremstyle{plain}
		\newaliascnt{theorem}{definition}	
		\newtheorem{theorem}[theorem]{Theorem}	
		\aliascntresetthe{theorem}	
		\newtheorem*{theorem*}{Theorem}	
		\newtheorem{conjecture}{Conjecture}

		\newtheorem*{conjecture*}{Conjecture}
	\theoremstyle{remark}
		\newtheorem*{remark}{Remark}
		\newtheorem*{example}{Example}

\usepackage{array} 
\usepackage{multicol} 
\usepackage[inline]{enumitem} 

\usepackage{microtype} 
	\microtypecontext{spacing=nonfrench}

\usepackage[backend=biber, bibstyle=authoryear, citestyle=alphabetic, hyperref, sorting=ynt, giveninits=true, maxbibnames=99, backref=false, doi=false]{biblatex}
	\DefineBibliographyStrings{english}{backrefpage = {see p.}, backrefpages = {see pp.}}
	\renewbibmacro{in:}{%
		\ifentrytype{article}{}{\printtext{\bibstring{in}\intitlepunct}}
	}
	\renewbibmacro*{doi+eprint+url}{
		\iftoggle{bbx:doi}{\printfield{doi}}{}
		\iftoggle{bbx:eprint}{\usebibmacro{eprint}}{}
		\iftoggle{bbx:url}{
			\iffieldundef{doi}{
				\iffieldundef{eprint}{
					\usebibmacro{url+urldate}
				}{}
			}{}
		}
	}
	\DeclareFieldFormat{labelalphawidth}{\mkbibbrackets{#1}}

	\defbibenvironment{bibliography}
		{\list
				{\printtext[labelalphawidth]{%
					\printfield{labelprefix}%
			\printfield{labelalpha}%
					\printfield{extraalpha}}}
				{\setlength{\labelwidth}{\labelalphawidth}%
				\setlength{\leftmargin}{\labelwidth}%
				\setlength{\labelsep}{\biblabelsep}%
				\addtolength{\leftmargin}{\labelsep}%
				\setlength{\itemsep}{\bibitemsep}%
				\setlength{\parsep}{\bibparsep}}%
				}
		{\endlist}
		{\item}

\usepackage[
	pdfusetitle,
	colorlinks=true, pdfborder={0 0 0},
	citecolor={dark-blue}, urlcolor={dark-blue},	linkcolor={dark-blue}
	]{hyperref}
			
\usepackage[numbered, open, openlevel=0]{bookmark}

\newcommand{\abs}[1]{\left| #1 \right|}
\newcommand{\norm}[1]{\left\lVert #1 \right\rVert}

\newcommand{\set}[1]{\left\{\ #1 \ \right\} } 
\newcommand{\suchthat}{\mathrel{}\middle|\mathrel{}} 

\newcommand{\spans}[1]{\operatorname{span}\left\{ #1 \right\}} 

\newcommand\restr{\raisebox{-.2ex}{\ensuremath \vert}} 

\DeclareMathOperator{\ad}{ad}
\DeclareMathOperator{\End}{End}
\DeclareMathOperator{\Conf}{Conf}
\DeclareMathOperator{\diag}{diag}
\newcommand{\id}{\textrm{id}} 

\newcommand{\HW}[1][]{{\mathop{\mathbf{HW}}{\hspace{-0.2em}}_{#1\, \/}}}
\newcommand{\dKW}[1][]{{\mathop{\mathbf{dKW}}{\hspace{-0.2em}}_{#1\, \/}}}
\newcommand{\EBE}[1][]{{\mathop{\mathbf{EBE}}{}_{#1\, \/}}}
\usepackage[left=2.5cm, right=2.5cm, top=3cm, bottom=3cm]{geometry}
\usepackage{setspace}
\let\savemathfrak\mathfrak 

\usepackage[ttscale=.85]{libertine} 
\usepackage[T1]{fontenc}
\usepackage[amsthm,upint]{libertinust1math} 

\let\mathfrak\savemathfrak 
\DeclareMathAlphabet\mathcal{OMS}{cmsy}{m}{n} 

\addtokomafont{section}{\large\scshape}
\addtokomafont{subsection}{\large\scshape}
\addtokomafont{paragraph}{\itshape\mdseries}

\RedeclareSectionCommand[
	afterskip=.2\baselineskip
]{section}

\RedeclareSectionCommand[
	afterskip=.2\baselineskip
]{subsection}

\title{Adiabatic Solutions of the Haydys-Witten Equations\\ and Symplectic Khovanov Homology}
\date{\today}

\begin{document}

\makeatletter
\renewcommand*{\thefootnote}{\fnsymbol{footnote}}

\thispagestyle{empty}

{

\setstretch{1.3}
\Large\bfseries\scshape
\@title

}

\vspace{1em}

Michael Bleher\footnotemark[2] \\
{\small\itshape
Institute for Mathematics, Heidelberg University, Im Neuenheimer Feld 205, Heidelberg, Germany.
}

\footnotetext[2]{mbleher@mathi.uni-heidelberg.de}
\footnotetext{\@date}

\renewcommand*{\thefootnote}{\arabic{footnote}}
\makeatother
\paragraph{Abstract.}
An influential conjecture by Witten states that there is an instanton Floer homology of four-manifolds with corners that in certain situations is isomorphic to Khovanov homology of a given knot $K$.
The Floer chain complex is generated by Nahm pole solutions of the Kapustin-Witten equations on $\mathbb{R}^3 \times \mathbb{R}^+_y$ with an additional monopole-like singular behaviour along the knot $K$ inside the three-dimensional boundary at $y=0$.
The Floer differential is given by counting solutions of the Haydys-Witten equations that interpolate between Kapustin-Witten solutions along an additional flow direction $\mathbb{R}_s$.
This article investigates solutions of a decoupled version of the Haydys-Witten equations on $\mathbb{R}_s \times \mathbb{R}^3 \times \mathbb{R}^+_y$, which in contrast to the full equations exhibit a Hermitian Yang-Mills structure and can be viewed as a lift of the extended Bogomolny equations (EBE) from three to five dimensions.
Inspired by Gaiotto-Witten's approach of adiabatically braiding EBE-solutions to obtain generators of the Floer homology, we propose that there is an equivalence between adiabatic solutions of the decoupled Haydys-Witten equations and non-vertical paths in the moduli space of EBE-solutions fibered over the space of monopole positions.
Moreover, we argue that the Grothendieck-Springer resolution of the Lie algebra of the gauge group provides a finite-dimensional model of this moduli space of monopole solutions.
These considerations suggest an intriguing similarity between Haydys-Witten instanton Floer homology and symplectic Khovanov homology and provide a novel approach towards a proof of Witten's gauge-theoretic interpretations of Khovanov homology.

\section{Introduction}
Let $M^5 = C \times \Sigma \times \mathbb{R}^+_y$, where $C$ and $\Sigma$ are Riemann surfaces, and assume $M^5$ is equipped with a product metric $g$ and the non-vanishing unit vector field $v=\partial_y$.
Write $\eta = g(v,\cdot)$ and observe that $\ker \eta$ coincides with the tangent space of $C\times \Sigma$.
Throughout, we let $J$ be the almost complex structure on $\ker \eta$ that is induced by the complex structures on $C$ and $\Sigma$, respectively.
We consider a principal $G$-bundle $E\to M^5$ for $G=SU(N)$ and denote by $\ad E$ its adjoint bundle.

In this article we investigate the decoupled Haydys-Witten (dHW) equations on $M^5$ with respect to $v$ and $J$ introduced in \cite{Bleher2023b}.
These are equations for a pair of connection $A \in \mathcal{A}(E)$ and Haydys' self-dual two-form $B\in\Omega^2_{v,+}(M^5, \ad E)$ (see \autoref{sec:khovanov-hermitian-yang-mills}).
The almost complex structure $J$ lifts to a map on $\Omega^2_{v,+}(M^5)$ with eigenvalues $\pm 1$.
The decoupled Haydys-Witten equations (dHW) are defined by
\begin{align}\label{eq:khovanov-decoupled-Haydys-Witten-equations}
	\tfrac{1+J}{2} \left( \sigma(B,B) + \nabla^A_v B \right) &= F_A^+ , &
	\tfrac{1-J}{2} \left( \sigma(B,B) + \nabla^A_v B \right) &= 0 , \\
	\delta_A^+ \tfrac{1+J}{2} B &= \imath_v F_A , &
	\delta_A^+ \tfrac{1-J}{2} B &= 0 .
\end{align}
In fact, we will mainly consider the decoupled Kapustin-Witten (dKW) equations, which arise from the dHW-equations as the $\mathbb{R}_s$-invariant reduction on five-manifolds of the form $M^5 = \mathbb{R}_s \times W^4$.
This reduction works in complete analogy to how the full Kapustin-Witten equations arise as the $\mathbb{R}_s$-invariant reduction of the Haydys-Witten equations and depends on the (constant) angle $\theta$ between $v$ and $\partial_s$ \cite{Bleher2024b}.

To fix notation for this reduction, let $w$ be the unit-norm vector field obtained from the projection of $v$ to $TW^4$, and let $v^\perp$ be the unit-norm vector field obtained from the subsequent projection of $w$ to $\ker \eta$.
We then write the connection as $A = A_s ds + \tilde A$ and reinterpret the $A_s$ component together with the three components of $B$ as a one-form $\phi = A_s w^\flat + \imath_{v^\perp} B$, where $w^\flat$ is the one-form dual to $w$.
The full Kapustin-Witten equations are a one-parameter family of equations for $(\tilde A, \phi) \in \mathcal{A}(E\vert_{W^4}) \times \Omega^1(W^4, \ad E\vert_{W^4})$ that depend on the angle $\theta$ \cite{Kapustin2007,Gagliardo2012}.

The main difference in the decoupled equations arises from the additional structure provided by the almost complex structure $J$ and its interplay with $v$.
Observe that there is a unique decomposition of the tangent bundle $TW^4$ into $J$-invariant subbundles $TW^4 = W_1 \oplus W_2$, given by $W_1 = \operatorname{span}\{ w, Jw \}$ and its orthogonal complement $W_2$.
This decomposition provides a splitting of the one-form $\phi = \varphi_1 + \varphi_2$ into components $\varphi_i \in \Omega^1(W^4, \ad E\vert_{W_i})$, $i=1,2$.
The decoupled Kapustin-Witten equations are then the following equations for the triple $(\tilde{A}, \varphi_1, \varphi_2)$.
\begin{align}\label{eq:khovanov-decoupled-Kapustin-Witten-equations}
	F_{\tilde A} - [\varphi_1 \wedge \varphi_1] - [\varphi_2 \wedge \varphi_2] + \cot\theta d_{\tilde A} \varphi_1 - \csc\theta \star_4 d_{\tilde A} \varphi_1 &= 0, \\
	-[\varphi_1 \wedge \varphi_2] + \cot\theta d_{\tilde A} \varphi_2 - \csc\theta \star_4 d_{\tilde A} \varphi_2 &= 0, \\
	d_{\tilde A} \star_4 \varphi_1 = d_{\tilde A} \star_4 \varphi_2 &= 0.
\end{align}

We are here interested in solutions of these decoupled equations in the context of Witten's gauge theoretic approach to Khovanov homology \cite{Witten2011}.
In this context, one constructs a Floer cochain complex from solutions of the Kapustin-Witten equations on a four-manifold $W^4 = X^3 \times \mathbb{R}^+_y$, subject to certain singular boundary conditions with monopole-like behaviour along a knot $K \subset \partial W^4 = X^3$.
Its cohomology with respect to the Floer differential, which counts the number of solutions of the (full) Haydys-Witten equations on $M^5 = \mathbb{R}_s \times W^4$ that interpolate between the Kapustin-Witten solutions at $s\to\pm\infty$, is expected to be a topological invariant that we shall call Haydys-Witten Floer homology.
Witten conjectures that for $X^3 = S^3$ or $\mathbb{R}^3$, this topological invariant coincides with Khovanov homology, for a more detailed description see e.g. \cite{Bleher2024b}.

In a highly influential article Gaiotto and Witten investigated Witten's conjecture by analysing the dimensional reduction of the Haydys-Witten and Kapustin-Witten equations to the extended Bogomolny equations (EBE) on three-manifolds of the form $\Sigma\times\mathbb R^+_y$ \cite{Gaiotto2012a}.
In contrast to the (full) Haydys-Witten and Kapustin-Witten equations, the EBE exhibit a Hermitian Yang-Mills structure and on that basis Gaiotto and Witten conjectured various Kobayashi-Hitchin type correspondences, relating EBE solutions with Nahm pole boundary conditions and knot singularities near a divisor $D\subset\Sigma\times\{0\}$ to suitable Higgs bundle data over $\Sigma$.

He and Mazzeo proved such Kobayashi-Hitchin correspondences for solutions of the EBE in terms of Higgs bundles \cite{He2019c,He2020b} and for a twisted version of the EBE in terms of Beilinson-Drinfeld opers \cite{He2019b}.
Various other works have investigated additional variations of the correspondence: e.g. Dimakis considered the case of nilpotent Higgs fields in \cite{Dimakis2022a}, while Sun considered the case of 'real-symmetry breaking' \cite{Sun2023}.
More recently, Dimakis refined the original Kobayashi-Hitchin type correspondence by identifying the moduli space of EBE solutions with a holomorphic Lagrangian inside the Higgs-bundle moduli space \cite{Dimakis2024a}.

Beyond the investigations of the EBE, Gaiotto and Witten proposed an `adiabatic braiding' procedure in which one transports EBE solutions along the direction the knot extends to produce Kapustin-Witten solutions.
They showed that the construction yields a braid-group action on the Poincar\'e polynomial of the Floer complex via the Jones representation of Virasoro conformal blocks, yet an explicit lift to Floer homology remains open.
He and Mazzeo proved that such a procedure works for $S^1$-invariant knots and in that way were able to lift their classification of EBE-solutions to a classification of $S^1$-invariant Kapustin-Witten solutions \cite{He2019a}.

The present work takes first steps toward a homological realisation of the Gaiotto-Witten programme by studying adiabatically braided solutions of the decoupled Haydys-Witten equations and relating them to non-vertical paths in the EBE moduli space.
Since the decoupled equations have the same Hermitian Yang-Mills structure as the EBE and contain them as a subset, they can be more easily linked to Higgs bundle data than the full equations.
Moreover, the vanishing results in \cite{Bleher2023a,Bleher2023b} suggest that, on nice enough manifolds, every Haydys-Witten or Kapustin-Witten solution is already a solution of the decoupled systems~\eqref{eq:khovanov-decoupled-Haydys-Witten-equations}~and~\eqref{eq:khovanov-decoupled-Kapustin-Witten-equations}, such that the adiabatic solutions may capture the full Haydys-Witten Floer homology.

In a parallel research direction, inspired by Haydys' original article \cite{Haydys2015} and ideas of Gaiotto-Moore-Witten \cite{Gaiotto2015b}, Wang developed a Landau-Ginzburg framework for Seiberg-Witten monopole Floer homology \cite{Wang2020a,Wang2020b,Wang2020c}.
In certain cases, this framework is related to Haydys-Witten instanton Floer theory and thus provides a monopole analogue of Witten's gauge-theoretic conjecture \cite{Wang2022a}.

It should also be noted that Gaiotto and Witten laid out an Atiyah-Floer type program to calculate the invariants associated to Haydys-Witten Floer theory, where instanton Floer theory is replaced by a Lagrangian intersection Floer theory.
For this, fix a Heegaard splitting $X^3 = H_1 \cup_{\Sigma} H_2$ and suppose that we stretch the metric transversely to $H_1\cap H_2$ such that the two handlebodies are joined by a long neck of the form $[-L,L]_t \times \Sigma$, $L>>1$.
Position the knot $K$ such that the portion of the knot in the long neck consists of a set of parallel straight lines $[-L, L]_t \times \{p_j\} $, intersecting $\Sigma \times \{0\}$ in a finite collection of points.
If $\mathcal{M}_\Sigma$ denotes the $\mathcal{G}_{\mathbb{C}}$ character variety of $\Sigma$, then the character varieties of the $H_i$ are Lagrangians $\mathcal{L}_1, \mathcal{L}_2 \subset \mathcal{M}_\Sigma$.
The moduli space of Kapustin-Witten solutions over $[-L,L]_t \times \Sigma\times \mathbb{R}^+_y$ that are invariant in the direction of $t$ provides a third Lagrangian $\mathcal{L}_3$.
In the absence of knots, the Atiyah-Floer conjecture states that Lagrangian intersection Floer homology of $\mathcal{L}_1$ and $\mathcal{L}_2$ is an invariant of $W^4$ and that this invariant coincides with the original instanton Floer cohomology, see \cite{Daemi2017, Abouzaid2020} for recent progress in this direction.
The effect of knots is included by counting instead holomorphic triangles that span between $\mathcal{L}_1$, $\mathcal{L}_2$, $\mathcal{L}_3$ in $\mathcal{M}_\Sigma$, which then conjecturally yields the coefficients of the Jones polynomial.
We refer to \cite{Gukov2017a} for more details and advances in this approach.

\begin{figure}[t]
	\centering
	\includegraphics[width=\textwidth, height=0.25\textheight, keepaspectratio]{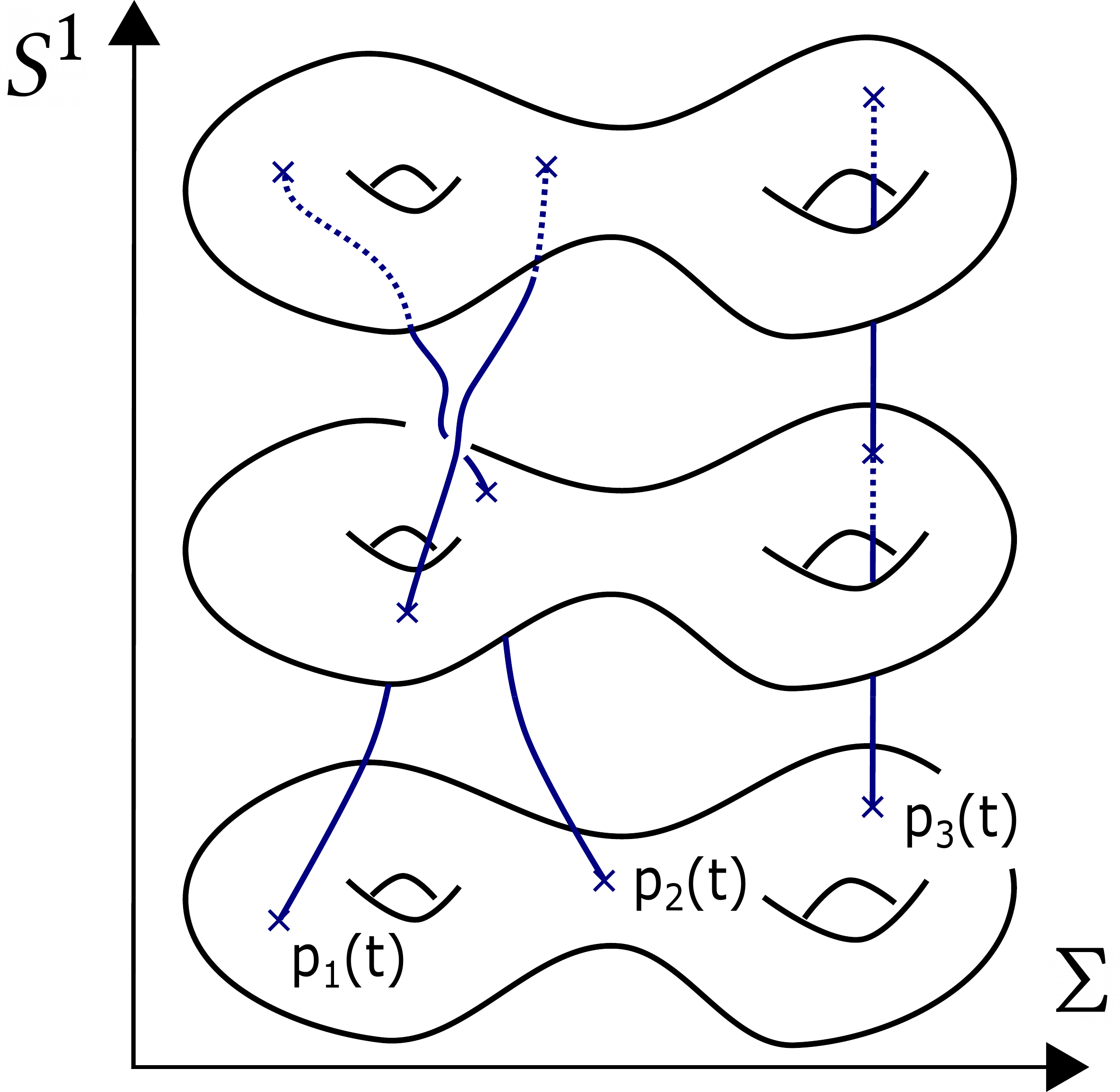}
	\caption{A general knot $K$ in the boundary of $W^4 = S^1_t \times \Sigma \times \mathbb{R}^+_y$ varies with time.
	The adiabatic approach can be viewed as stretching the size of $S^1_t$, such that at any given time $t$, the fields $(A,\phi)$ are well-approximated by a solution of the extended Bogomolny equations.}
	\label{fig:khovanov-general-braid}
\end{figure}

In the present work, however, we remain on the side of instanton Floer theory and investigate the adiabatic approach for solutions of the decoupled Haydys-Witten and Kapustin-Witten equations directly.
The adiabatic condition corresponds to the assumption that on a given slice $[-L,L]_t \times \Sigma \times \mathbb{R}^+_y$ the knot position varies only slightly in $t$ (cf. \autoref{fig:khovanov-general-braid}), such that one obtains an approximate Kapustin-Witten solution from a smooth family of EBE-solutions that ``remain in the ground state'' when one moves from $t=-L$ to $L$.
This means that Kapustin-Witten solutions should be related to certain well-behaved paths in the moduli space of EBE-solutions.

Since this is the key idea of the article, it deserves a more detailed explanation.
Consider the decoupled Kapustin-Witten (dKW) equations on $S^1_t\times \mathbb{C} \times \mathbb{R}^+_y$ and a knot of the form $K=\{ (t, p_a(t), 0) \}_{a=1, \ldots, k}$.
The collection of trajectories $\{p_a(t)\}_{a=1,\ldots, k}$ can be viewed as a loop $\beta: S^1_t \to \Conf_k \mathbb{C}$ in the configuration space of $k$ distinct, ordered points in $\mathbb{C}$.
Write $\mathcal{M}^{\mathrm{dKW}}_K$ for the moduli space of dKW-solutions subject to Nahm pole boundary conditions with knot singularities along $K$.
We can view $\mathcal{M}^{\mathrm{dKW}}_K$ as the fiber of a bundle $\mathcal{M}^{\mathrm{dKW}} \to \Omega \Conf_k \mathbb{C}$, where the fiber map sends each solution to the knot $K$ at which the solution exhibits a knot singularity.
There is an analogous fiber bundle for solutions of the extended Bogomolny equations $\mathcal{M}^{\mathrm{EBE}} \to \Conf_k \mathbb{C}$ that sends a solution to the position of the knot singularities in a temporal slice $\{t\} \times \mathbb{C} \times \mathbb{R}^+_y$, where they are given by a divisor $D = \{ p_a \}_{a=1,\ldots, k} \subset \Sigma$.

The adiabatic approach should be viewed as the statement that one expects there to be a bundle map of the form
\begin{equation}
	\label{eq:khovanov-bundle-map}
	\begin{tikzcd}
		\mathcal{M}^{\mathrm{dKW}}  \arrow[rr, dashed] \arrow[rd] &[-30pt]
		&[-30pt]
		\Omega \mathcal{M}^{\mathrm{EBE}} \arrow[dl]
		\\
		& 
		\Omega \Conf_k \Sigma &
		&
	\end{tikzcd}
\end{equation}
The following result by He and Mazzeo establishes the existence of such a map for the case of $S^1_t$-invariant Kapustin-Witten solutions.
\begin{theorem*}[\cite{He2019a}]
\label{thm:khovanov-He-Mazzeo-S1-invariant-classification}
Every solution of the EBE on $\Sigma\times\mathbb{R}_+$ with Nahm pole boundary condition and knot singularities at a divisor $D = \{z_a\}_{a=1,\ldots, k} \subset \Sigma$ lifts to an $S^1_t$-invariant solution of the KW-equations on $S^1_t \times \Sigma \times \mathbb{R}^+_y$ with knot singularities along the $S^1_t$-invariant knot $K= S^1_t \times D$.
Moreover, every $S^1_t$-invariant solution of the KW-equations is given by such a lift and there is a bijection between moduli spaces
\begin{align}
	\mathcal{M}^{\mathrm{KW}}_{S^1_t \times D} \to \mathcal{M}^{\mathrm{EBE}}_{D} .
\end{align}
\end{theorem*}
Note that, as dimensional reduction of the Kapustin-Witten equations, any EBE-solution immediately provides an $S^1_t$-invariant solution of the Kapustin-Witten equations.
The content of the theorem is that also the reverse is true: any solution for $S^1_t$-invariant knots is necessarily a lift of an EBE-solution -- or equivalently a lift of a constant loop in $\Omega \mathcal{M}^{\mathrm{EBE}}$.
This result also provides a restriction on the bundle map in~\eqref{eq:khovanov-bundle-map} if $K$ is not constant along $S^1_t$: the range must be the space of non-vertical or horizontal loops, which we denote by $\Omega_h \mathcal{M}^{\mathrm{EBE}}$.

The proof of He and Mazzeo's theorem is based on a Weitzenb\"ock formula that equates the full Kapustin-Witten equations with the EBE up to certain boundary terms.
A prerequisite for this Weitzenb\"ock formula is the assumption that the position of knot singularities is $S^1_t$-invariant.
This assumption will be dropped in the analysis presented here, such that the classification of solutions requires a richer structure.

Instead of the Weitzenb\"ock formula of He and Mazzeo, we rely on the Hermitian Yang-Mills structure\footnotemark of the decoupled Haydys-Witten equations~\eqref{eq:khovanov-decoupled-Haydys-Witten-equations}.
\footnotetext{%
In a sense, the Weitzenb\"ock formula of He and Mazzeo is replaced by the one of \cite{Bleher2023b} that establishes the decoupling of the equations on $M^5 = C \times \Sigma \times \mathbb{R}^+_y$.
}
This structure is most easily described in the $4\mathcal{D}$-formalism, where one uses the Haydys-Witten fields $(A,B)$ to define four $\ad E_{\mathbb{C}}$-valued differential operators $\mathcal{D}_\mu$, $\mu=0,1,2,3$ (see \autoref{sec:khovanov-hermitian-yang-mills}).
In terms of these operators, the decoupled Haydys-Witten equations are equivalent to
\begin{align}
	[ \mathcal{D}_\mu, \mathcal{D}_\nu] = 0,\ \mu,\nu = 0,1,2,3, \qquad \sum_{\mu=0}^3 [\mathcal{D}_\mu, \overline{\mathcal{D}_\mu}] = 0 .
\end{align}
Crucially, the set of equations on the left is invariant under complex gauge transformations $\mathcal{G}_{\mathbb{C}}$, while the remaining equation on the right can be viewed as a real moment map condition.
As a consequence, it's possible to first solve the easier $\mathcal{G}_{\mathbb{C}}$-invariant part of the equations and subsequently solve for a complex gauge transformation that solves the moment map condition.

In a first step, following the adiabatic approach of Gaiotto and Witten, we consider a situation where the knot $K$ is a small deformation of an $S^1$-invariant one $K^\prime$.
More precisely, we assume that the two knots are connected by a well-behaved isotopy $\beta_\bullet$, where $\beta_0 = K^\prime$ and $\beta_1 = K$.
Since the underlying physical theory is topological, such an isotopy can not have an effect on the number of solutions.
We then introduce an ansatz that solves the $\mathcal{G}_{\mathbb{C}}$-invariant part of the decoupled Haydys-Witten equations and exhibits a knot singularity along the knot trajectories determined by $\beta_q(t)$ at each point $q \in[0,1]$ of the isotopy.
Put differently, the isotopy provides a deformation from an initially $S^1$-invariant ansatz with knot singularities along the $S^1$-invariant knot $K^\prime$, to a $t$-dependent ansatz with knot singularities along the $t$-dependent knot $K$.
We propose that this isotopy ansatz provides a way to transport a solution of the decoupled Haydys-Witten equations with $S^1$-invariant knot to a solution with $t$-dependent knot.

A second key aspect of this work is a physically motivated reduction from the infinite-dimensional moduli space of Kapustin-Witten solutions (before gauge fixing) to a finite-dimensional model space that retains enough information to find solutions.
The model space in question is related to a partial Grothendieck-Springer resolution of $\mathfrak{sl}(kN)$ that is naturally fibered over the configuration space of $k$ points in $\mathbb{C}$.
Motivated by observations about the isotopy ansatz, we formulate \autoref{conj:lower-bound-for-KW-solns} stating that on $S^1_t \times \Sigma \times \mathbb{R}^+_y$, the number of intersection points of the Grothendieck-Springer fiber and its parallel transport along $S^1_t$ determines a lower bound for the number of solutions to the decoupled Kapustin-Witten equations.
Taking this line of argument to its logical conclusion leads to \autoref{conj:khovanov-HF-is-symp-Kh}, which claims that Haydys-Witten Floer theory is isomorphic to symplectic Khovanov-Rozansky homology as defined by Seidel, Smith and Manolescu \cite{Seidel2004, Manolescu2007}.
Notably, recent work by Tan et al. has independently derived a closely related correspondence~\cite[Sect.~9.5]{Er2023}.
Since symplectic Khovanov homology (associated to $G_{\mathbb{C}} = SL(2,\mathbb{C})$) is known to be isomorphic to a grading reduced version of Khovanov homology \cite{Abouzaid2019}, the arguments developed here provide a novel approach to prove Witten's conjecture, complementary to the Atiyah-Floer approach pursued in the Gaiotto-Witten program.
\vspace{\baselineskip}

\noindent
The article is arranged into two parts:

Sections~\ref{sec:khovanov-hermitian-yang-mills}~-~\ref{sec:khovanov-kobayashi-hitchin-correspondence} provide a review of the relevant background and we also use the opportunity to fix some notation used in the remainder of this article.
Specifically,
\autoref{sec:khovanov-hermitian-yang-mills} spells out the Hermitian Yang-Mills structure of the decoupled Haydys-Witten equations;
\autoref{sec:khovanov-orbits-and-slices} provides some notation regarding Lie algebras and adjoint orbits;
\autoref{sec:khovanov-Nahm-pole-boundary-condition} introduces the Nahm pole boundary conditions with knot singularities in a form that is adjusted to the Hermitian Yang-Mills structure;
and \autoref{sec:khovanov-kobayashi-hitchin-correspondence} summarizes the results of He and Mazzeo for solutions of EBE-solutions \cite{He2019c, He2020b}, highlighting certain aspects that will have close analogues in subsequent discussions for solutions of the decoupled Haydys-Witten and Kapustin-Witten equations.

Sections~\ref{sec:khovanov-isotopy-ansatz}~-~\ref{sec:khovanov-homology} formalize the adiabatic approach:
We introduce the isotopy ansatz and an associated expansion of the decoupled Haydys-Witten equations in~\autoref{sec:khovanov-isotopy-ansatz} and subsequently suggest in \autoref{sec:khovanov-method-of-continuity} a strategy to obtain solutions of the decoupled Kapustin-Witten equations for any null-isotopic single-stranded knot by use of a continuity argument.
In \autoref{sec:khovanov-comoving-higgs-bundles} we explain the resulting relation between solutions of the decoupled Haydys-Witten equations and paths in the moduli space of Higgs bundles $(\mathcal{E},\varphi)$ that are equipped with the additional structure of a distinguished line subbundle $L$.
Based on physical intuition, we propose that for our purposes the Grothendieck-Springer fibration can be used to model the moduli space of triples $(\mathcal{E},\varphi,L)$ in \autoref{sec:khovanov-effective-triples-and-monopoles} and using these insights, we formulate \autoref{conj:lower-bound-for-KW-solns} that provides a lower bound for the number of Kapustin-Witten solutions.
In \autoref{sec:khovanov-braids-to-knots} we explain that one naturally obtains Lagrangian submanifolds of the fibers when the $S^1_t$-factor is decompactified and replaced by $\mathbb{R}_t$ by identifying compact knots with braid closures.
Finally, \autoref{sec:khovanov-homology} incorporates Haydys-Witten instantons into the setting, which leads to \autoref{conj:khovanov-HF-is-symp-Kh} that Haydys-Witten instanton Floer homology coincides with symplectic Khovanov homology.
\paragraph{Acknowledgements}
The author acknowledges support by the Heidelberg Graduate School for Fundamental Physics (HGSFP) during the initial phase of this research.
This work is funded by the Deutsche Forschungsgemeinschaft (DFG, German Research Foundation) under Germany's Excellence Strategy EXC 2181/1 - 390900948 (the Heidelberg STRUCTURES Excellence Cluster).
\section{Hermitian Yang-Mills Structure of the Decoupled Haydys- and Kapustin-Witten Equations}
\label{sec:khovanov-hermitian-yang-mills}

Let $M^5 = C \times \Sigma \times \mathbb{R}^+_y$, where $C$ and $\Sigma$ are Riemann surfaces, equipped with a product metric $g$ and fix the non-vanishing unit vector field $v=\partial_y$ with dual one-form $\eta=dy$.
In this situation $\ker \eta$ coincides with the tangent space of $C\times \Sigma$.
Let $J$ be the almost complex structure on $\ker \eta$ that is induced by the complex structures on $C$ and $\Sigma$.

In the context of Haydys-Witten Floer theory, we always assume that $C$ contains the non-compact flow direction and correspondingly is either $C\simeq \mathbb{R}_s \times \mathbb{R}_t$ or $\mathbb{R}_s \times S^1_t$ with the corresponding standard complex structure.
In contrast, $\Sigma$ might generally be an arbitrary Riemann surface.
In the end, we will be most interested in the special case $\Sigma=\mathbb{C}$, because in that case Haydys-Witten Floer theory can be related to Khovanov homology.
We let $(w, z)$ denote holomorphic coordinates on $C\times \Sigma$ and will also write $w = s+it$ and $z=x^2+i x^3$ in terms of real coordinates.

Consider a $G=SU(N)$-principal bundle $E$ over $M^5$ and let $G_{\mathbb{C}}$ and $E_{\mathbb{C}}$ denote the corresponding complexifications.
Let $A$ be a gauge connection on $E$ and $B\in \Omega^2_{v,+}(M^5, \ad E)$ an element of  Haydys' self-dual two-forms with respect to $v=\partial_y$ \cite{Haydys2015}.
In holomorphic coordinates this means that $B$ is given by
\begin{align}
	B = i \phi_1 ( dw \wedge d\bar{w} + dz \wedge d\bar{z}) + \varphi\, dw \wedge dz + \overline{\varphi}\, d\bar{w} \wedge d\bar{z} .
\end{align}
In real coordinates, this corresponds to $B = \sum_{a=1}^3 \phi_a (dx^0 \wedge dx^a + \frac{1}{2} \epsilon_{abc} dx^b\wedge dx^c)$, where components are related by $\varphi = \phi_2 - i \phi_3$, $\overline{\varphi} = \phi_2 + i \phi_3$.

Introduce the following differential operators $\mathcal{D}_\mu$ that act on sections of $\ad E_{\mathbb{C}}$.
\begin{align}
\label{eq:khovanov-definition-four-Ds}
	\mathcal{D}_0 &= 2 \nabla^A_{\bar w} = \nabla^A_0 + i\nabla^A_1 , &
	\mathcal{D}_1 &= 2 \nabla^A_{\bar z} = \nabla^A_2 + i\nabla^A_3 , \\
	\mathcal{D}_2 &= \nabla^A_y - i [\phi_1, \cdot] , &
	\mathcal{D}_3 &= [\varphi , \cdot] = [\phi_2, \cdot] - i [\phi_3,\cdot] .
\end{align}
The complex structure of $\ad E_{\mathbb{C}}$ induces a complex conjugation that we will denote by $\overline{\mathcal{D}_\mu}$.
Furthermore, there is an action of $G_{\mathbb{C}}$-valued gauge transformations $g(x) \in \mathcal{G}_{\mathbb{C}}(M^5)$ by conjugation $\mathcal{D}_\mu \mapsto g(x)^{-1} \mathcal{D}_\mu g(x)$.

The full Haydys-Witten equations \cite{Witten2011, Haydys2015} are the equations
\begin{align}
	\label{eq:khovanov-haydys-witten-4D}
	[\overline{\mathcal{D}_0}, \overline{\mathcal{D}_i}] - \tfrac{1}{2} \epsilon_{ijk} [\mathcal{D}_j , \mathcal{D}_k] = 0 ,\ i,j,k=1,2,3,\
	\qquad
	\sum_{\mu=0}^3 [\overline{\mathcal{D}_\mu}, \mathcal{D}_\mu] = 0 ,
\end{align}
while the decoupled Haydys-Witten equations correspond to the specialization \cite{Bleher2023b}
\begin{align}
	\label{eq:khovanov-decoupled-haydys-witten-4D}
	[\mathcal{D}_\mu ,\mathcal{D}_\nu] = 0,\ \mu,\nu = 0,1,2,3\, ,\
	\qquad 
	\sum_{\mu=0}^3 [ \overline{\mathcal{D}_\mu} , \mathcal{D}_\mu] = 0 .
\end{align}
As $\mathbb{R}_s$-invariant reduction, the decoupled Kapustin-Witten equations are given by exactly the same formula but with $\mathcal{D}_0 = [A_s, \cdot] + i \nabla^A_1$, where $A_s$ must be reinterpreted as the $\phi_y$ component of the four-dimensional one-form $\phi$.
Similarly, the decoupled Haydys-Witten equations on $C\times \Sigma \times \mathbb{R}^+_y$ reduce to the three-dimensional extended Bogomolny equations (EBE) on $\Sigma \times \mathbb{R}^+_y$ when the five-dimensional fields are independent of $C$, since for regular solutions a vanishing result implies that $\mathcal{D}_0 \equiv 0$.

Let us comment further on the Hermitian Yang-Mills structure of the decoupled equations.
For this, consider the submanifold $C\times\Sigma\times \{y\}$ for some fixed $y$.
The structure we have described above becomes that of a K\"ahler manifold $(C\times \Sigma, \omega)$, with K\"ahler form $\omega = g(J\cdot, \cdot)$, together with a complex vector bundle $\ad E_{\mathbb{C}}$.
This vector bundle is equipped with a Hermitian metric $h$, with respect to which $\mathcal{A}$ is Hermitian.

The four operators $\mathcal{D}_\mu$ and their complex conjugates can be viewed as holomorphic and anti-holomorphic components of a \emph{complexified} covariant derivative associated to the Hermitian connection $\mathcal{A}$ on $\ad E_{\mathbb{C}}$.
By this we mean the $\mathbb{C}$-linear map $\nabla^{\mathcal{A}}: \Gamma(\ad E_{\mathbb{C}})\to \Gamma(T^*_{\mathbb{C}} M \otimes \ad E_{\mathbb{C}})$ that is locally given by $\nabla^{\mathcal{A}}_{\partial_\mu} = \mathcal{D}_\mu$ and $\nabla^{\mathcal{A}}_{\overline{\partial_\mu}} = \overline{\mathcal{D}}_\mu$.
Here we denote by $\overline{\partial_\mu} = \hat J \partial_\mu$ a local orthonormal frame of $T_{\mathbb{C}} M$ with respect to the standard complex structure $\hat J$ on the complexified tangent bundle of $M$.
The covariant derivative has the property that it vanishes automatically in the direction of the vector field $v=\partial_y$, i.e. $\mathcal{D}_y = 0$.
Let $F_{\mathcal{A}} \in \Omega_{\mathbb{C}}^2(M^5, \ad E_{\mathbb{C}})$ be its curvature two-form, defined by
\begin{align}
	(F_{\mathcal{A}})_{\mu\nu}\; s = [\mathcal{D}_\mu, \mathcal{D}_\nu] s .
\end{align}
Denote by $F_{\mathcal{A}}^{p,q}$ the $(p,q)$-part of the field strength.
Since $\mathcal{D}_y = 0$, the field strength $F_{\mathcal{A}}$ is an element of the subbundle $\Omega^2_{v,-}(M^5, \ad E_{\mathbb{C}}) \oplus \Omega^2_{v,+}(M^5, \ad E_{\mathbb{C}})$, where the map $T_\eta := \star_5 ( \eta \wedge \cdot )$ acts on the summands with eigenvalues $\pm 1$, respectively.
The anti-self-dual anti-holomorphic part of $F_{\mathcal{A}}$ is then given by
\begin{align}
  \begin{multlined}
    \frac{1}{2} \left( F_{\mathcal{A}}^{2,0} - T_\eta \hat{J} F_{\mathcal{A}}^{0,2} \right) \\
      = \sum \big( (F_{\mathcal{A}})_{0i} - \tfrac{1}{2} \epsilon_{ijk}\; (\overline{F_{\mathcal{A}}})_{jk}\ \big) \big( dx^0\wedge dx^i + \tfrac{1}{2} \epsilon_{ijk}\; d x^j\wedge d x^k \big).
  \end{multlined}
\end{align}
Furthermore, there is an inner product $\Lambda_\omega: \Omega_{\mathbb{C}}^{1,1}(C\times \Sigma) \to \Omega_{\mathbb{C}}^0(C\times \Sigma)$, induced by the K\"ahler form $\omega = g(J\cdot, \cdot)$ and normalized such that in coordinates and if the metric on $C\times \Sigma$ is flat $\omega = i/2 ( dx^0 \wedge d \bar{x}^0 + \ldots + dx^3 \wedge d \bar{x}^3)$.
Application of $\Lambda_\omega$ to $F_\mathcal{A}$ corresponds to the trace of its mixed part $[\overline{\mathcal{D}_\mu}, \mathcal{D}_\nu]$.
With this the Haydys-Witten equations~\eqref{eq:khovanov-haydys-witten-4D} become anti-holomorphic anti-self-duality equations
\begin{align}
	F_{\mathcal{A}}^{2,0} - T_\eta \hat{J} F_{\mathcal{A}}^{0,2} = 0, \quad
	\Lambda_\omega F_\mathcal{A} = 0,
	\label{eq:khovanov-Haydys-Witten-equations}
\end{align}
while the decoupled Haydys-Witten equations~\eqref{eq:khovanov-decoupled-haydys-witten-4D} are equivalent to the Hermitian Yang-Mills equations
\begin{align}
	F_{\mathcal{A}}^{2,0} =0, \quad 
	\Lambda_\omega F_{\mathcal{A}} = 0.
	\label{eq:khovanov-hermitian-yang-mills}
\end{align}
The first equation, $F_{\mathcal{A}}^{2,0} = 0$, is invariant under complex gauge transformations $g \in \mathcal{G}_{\mathbb{C}}(M^5)$, while the equation $\Lambda_\omega F_{\mathcal{A}} = 0$ is only invariant under the subgroup of unitary gauge transformations with respect to the metric $h$.
More explicitly: those gauge transformation that satisfy $\overline{g} h g = h$.
Accordingly, the second equation describes a real moment map condition.
This extends the Hermitian-Yang-Mills structure previously observed for the EBE to the decoupled Haydys-Witten equations.

The work of Donaldson \cite{Donaldson1985, Donaldson1987} and Uhlenbeck-Yau \cite{Uhlenbeck1986} shows that the geometric data of solutions to the $G_{\mathbb{C}}$-invariant equations play an important role in understanding the solutions of the full equations.
The main underlying idea is that one can first solve the easier $\mathcal{G}_{\mathbb{C}}$-invariant equations and subsequently try to find a complex gauge connection such that the solution satisfies the remaining real moment map condition.
Indeed, the model solutions for the Nahm pole boundary conditions are constructed in this way, and this is what will be discussed in the next section.


\section{Adjoint Orbits and Slodowy Slices}
\label{sec:khovanov-orbits-and-slices}

This section introduces the relevant notation and certain standard constructions for the Lie algebra $\mathfrak{sl}(N,\mathbb{C})$ that will be used in the rest of this article, for a more detailed discussion see for example \cite{Collingwood1993}.
Throughout, we choose and fix a Cartan subalgebra $\mathfrak{h}$ and a Chevalley basis $\{H_i, E_i^\pm\}_{i\in\{1,\ldots, N-1\}}$ of $\mathfrak{sl}(N,\mathbb{C})$ with Cartan matrix $A_{ij}$.
The Lie bracket satisfies
\begin{align}
	[H_i, H_j] = 0 ,\qquad
	[H_i, E_j^\pm] = \pm A_{ji} E_j^\pm , \qquad
	[E_i^+, E_j^-] = \delta_{ij} H_i.
\end{align}

An element $X \in \mathfrak{sl}(N,\mathbb{C})$ is called regular if the dimension of its centralizer $Z_{\mathfrak{sl}(N,\mathbb{C})}(X) = \{ Y \in \mathfrak{sl}(N,\mathbb{C}) | [Y,X] = 0 \}$ is minimal, i.e. it is equal to the dimension of the Cartan subalgebra $\mathfrak{h}$.
An element $E\in\mathfrak{sl}(N,\mathbb{C})$ is nilpotent if there is a positive integer such that $(\ad_E)^n = 0$.
Let $\pi_1, \ldots, \pi_s$ be a collection of positive integers that satisfy $\pi_1 + \ldots + \pi_s = N$.
Denoting by $J_{\pi_i}(\lambda)$ a Jordan block of size $\pi_i$ with eigenvalue $\lambda$, the Jordan normal form of a nilpotent element is given by
\begin{align}
	E_\pi = \begin{pmatrix} J_{\pi_1}(0) & & \\ & \ddots & \\ & & J_{\pi_s}(0) \end{pmatrix}.
\end{align}
More generally, we use the notation $\pi = [\pi_1{}^{\nu_1} \ldots \pi_s{}^{\nu_s}]$ for partitions of $N$, where $\nu_i$ denote multiplicities such that $N = \nu_1 \pi_1 + \ldots + \nu_s\pi_s$, and entries are ordered according to $\pi_1 \geq \pi_2 \geq \ldots \pi_s$ .
The Jordan normal form of a regular nilpotent element is associated to the partition $\pi=[N]$, while the Jordan normal form of $0$ corresponds to $\pi = [1{}^N]$.

The orbit of an element $X\in\mathfrak{sl}(N,\mathbb{C})$ under the adjoint action of $SL(N,\mathbb{C})$ is denoted by
\begin{align}
	\mathcal{O}(X) := \set{ g X g^{-1} \suchthat g \in SL(N,\mathbb{C}) }.
\end{align}
We write $\mathcal{O}_\pi = \mathcal{O}(E_\pi)$ and, in fact, any orbit of nilpotent elements is of that form: for any nilpotent element $E$ the associated partition $\pi$ is determined by its Jordan normal form.

The orbit $\mathcal{O}_{[N]}$ corresponds to the adjoint orbit of regular nilpotent elements and will also be denoted by $\mathcal{O}_{\mathrm{reg}}$.
The closure of the regular nilpotent orbit $\mathcal{N} := \overline{\mathcal{O}_{\mathrm{reg}}}$ is called the nilpotent cone of $\mathfrak{sl}(N,\mathbb{C})$.
The nilpotent cone is an algebraic variety that contains all nilpotent orbits of $\mathfrak{sl}(N,\mathbb{C})$ as singular loci.

Nilpotent orbits come with a partial order, defined by setting $\mathcal{O}_\pi \leq \mathcal{O}_\rho$ if $\mathcal{O}_\pi \subseteq \overline{\mathcal{O}_\rho}$.
This partial order is equivalent to the dominance order on the set of partitions of $N$, where $\pi \leq \rho$ if and only if $\pi_1 + \ldots + \pi_k \leq \rho_1 + \ldots + \rho_k$ for all $k$.

The Jacobson-Morozov theorem states that for any non-zero nilpotent element $E$ there exists an $\mathfrak{sl}_2$-triple $(E,H,F)$, i.e. elements that satisfy the $\mathfrak{sl}(2,\mathbb{C})$ commutation relations $[H,E] = 2 E$, $[E,F] = H$, $[H, F] = -2 F$.
Moreover, $H$ and $F$ are unique up to conjugation by elements of the centralizer $Z_{SL(N,\mathbb{C})}(E)=\{g \in SL(N,\mathbb{C}) | g E g^{-1} = E \}$.

Let $\pi$ be a partition of $N$ and fix a nilpotent element $E\in\mathcal{O}_{\pi}$.
Choose a completion to an $\mathfrak{sl}_2$-triple $(E,H,F)$.
The affine subspace of $\mathfrak{sl}(N,\mathbb{C})$ defined by 
\begin{align}
	\mathcal{S}_{E} := E + \ker \ad_{F},
\end{align}
is called the Slodowy slice to $\mathcal{O}_{\pi}$ at $E$.
Slodowy slices are transversal slices in $\mathfrak{sl}(N,\mathbb{C})$, meaning that they have transverse intersections with all adjoint orbits in $\mathfrak{sl}(N, \mathbb{C})$.
We will also write $\mathcal{S}_\pi$ for the Slodowy slice to $\mathcal{O}_{\pi}$ at our favourite element of the orbit, the Jordan normal form $E_\pi$.


\section{The Hermitian Version of the Nahm Pole Boundary Conditions}
\label{sec:khovanov-Nahm-pole-boundary-condition}

The Nahm pole boundary conditions are modeled on certain singular solutions of Nahm's equations over $\mathbb{R}^+_y$ and, in the presence of knots, on monopole solutions of the EBE over $\mathbb{C} \times \mathbb{R}^+_y$ \cite{Witten2011, Mazzeo2014, Mazzeo2017}.
Both of these equations are dimensional reductions\footnotemark of the decoupled Haydys-Witten equations~\eqref{eq:khovanov-decoupled-haydys-witten-4D}.
\footnotetext{%
See \cite{Bleher2024b} for a comprehensive discussion of the dimensional reductions of the Haydys-Witten equations.
}
The EBE arise for $\mathcal{D}_0=0$, while Nahm's equations correspond to the case $\mathcal{D}_0 = \mathcal{D}_1 = 0$.
Note that both equations retain the Hermitian Yang-Mills structure of the decoupled Haydys-Witten equations.

In this section we first provide a short review of the derivation of the model solutions.
This was described for $SU(2)$ by Witten and later for $SU(N)$ by Mikhaylov \cite{Witten2011, Mikhaylov2012}.
In \autoref{sec:khovanov-isotopy-ansatz} we will extend the original ansatz of Witten and Mikhaylov presented here to the situation of the decoupled Haydys-Witten equations with $\mathcal{D}_0 \neq 0$.
The section concludes with a definition of the Nahm pole boundary conditions when viewed as a condition on a complex gauge transformation.
This reformulation was originally described very similarly by He and Mazzeo in \cite{He2019c, He2020b}.

For the rest of this section assume that $\Sigma = \mathbb{C} $ with holomorphic coordinate $z=x^2+ix^3$.
We will also use polar coordinates $z = r e^{i\vartheta}$ on $\mathbb{C}$ and (hemi-)spherical coordinates $(R,\psi,\vartheta)$ on $\mathbb{C} \times \mathbb{R}^+_y$, where $R = r^2 + y^2$, $\cos\psi = \frac{y}{R}$, and $\vartheta$ remains the azimuthal angle in the complex plane.
Note that in spherical coordinates the boundary $y=0$ corresponds to points with $\psi = \pi/2$.

The Hermitian Yang-Mills structure of the decoupled Haydys-Witten equations described in \autoref{sec:khovanov-hermitian-yang-mills} suggests a way to solve the equations \cite{Witten2011, Mikhaylov2012, Gaiotto2012a}:
Following the ideas of Donaldson-Uhlenbeck-Yau \cite{Donaldson1985, Uhlenbeck1986, Donaldson1987}, one starts from holomorphic data that satisfies the $\mathcal{G}_{\mathbb{C}}$-invariant equations $[\mathcal{D}_i,\mathcal{D}_j] = 0$ and then determines a gauge transformation $g\in \mathcal{G}_{\mathbb{C}}(\mathbb{C}\times \mathbb{R}^+_y)$ that solves the moment map condition $\sum_{i=1}^3 [\overline{\mathcal{D}_i}, \mathcal{D}_i] = 0$.
We start with a description of Nahm pole solutions and afterwards discuss monopole solutions, which additionally incorporate knot singularities.

\subsection{Nahm Pole Solutions}
Let $E \in \mathcal{N}$ be a nilpotent element and consider as an initial ansatz
\begin{align}
	A^0 = 0, \qquad 	
	\varphi^0 = E, \qquad
	\phi_1^0 = 0.
	\label{eq:khovanov-nahm-pole-singularity-ansatz}
\end{align}
Using the definitions in~\eqref{eq:khovanov-definition-four-Ds}, this field configuration clearly satisfies the $G_{\mathbb{C}}$-invariant part of Nahm's equations $[\mathcal{D}_2, \mathcal{D}_3] = 0$.
Note that, since $\varphi = \phi_2+i\phi_3$, the complex conjugate $\overline{\varphi}^0 = \phi_2 -i \phi_3 =: F$ and this determines a unique $\mathfrak{sl}_2$-triple $(E,H,F)$.

We now ask for a complex gauge transformation $g_0 \in \mathcal{G}_{\mathbb{C}}(\mathbb{R}^+_y)$ that maps the fields of this initial ansatz to a solution of the real moment map equation $\sum_{i=2}^3 [\overline{\mathcal{D}}_i, \mathcal{D}_i] = 0$.
Since the unitary part of the gauge transformation drops out, we can assume that $g_0$ takes values in $\exp i\mathfrak{g}$ with respect to the Cartan decomposition $G_{\mathbb{C}} = G \cdot \exp i\mathfrak{g}$.
Writing $g_0 = \exp \psi$, assuming that $\psi \in i\mathfrak{h}$ and only depends on $y$, and plugging the transformed operator $g_0 \mathcal{D}_i g_0^{-1}$ into the moment map equation leads to
\begin{align}
	\label{eq:khovanov-moment-map-after-gauge-trafo-1d}
	\partial_y^2 \psi + \frac{1}{2} [\overline{\varphi}^0, e^{2\psi} \varphi^0 e^{-2\psi}] &= 0.
\end{align}
This equation has a simple solution\footnotemark, given by
\footnotetext{%
Use Lie's expansion formula (also attributed to Campbell and Hadamard): $e^{X} Y e^{-X} = \sum [X,Y]_k/k!$, where $[X,Y]_k = [X,[X,Y]_{k-1}]$ and $[X,Y]_0 = Y$.
}
\begin{align}
	g_0 = \exp(-\log y\ H).
	\label{eq:khovanov-nahm-pole-singularity-transformation}
\end{align}
The action of $g_0$ on $\mathcal{D}_\mu$ transforms the initial ansatz~\eqref{eq:khovanov-nahm-pole-singularity-ansatz} into the desired Nahm pole solution
\begin{align}
	A = 0, \qquad
	\varphi = \frac{E}{y}, \qquad
	\phi_1 = \frac{H}{y}.
\end{align}
Observe that the choice of nilpotent element $E \in \mathcal{N}$ in the initial ansatz~\eqref{eq:khovanov-nahm-pole-singularity-ansatz} uniquely determines the Nahm pole solution.
We call the solution a regular Nahm pole if $E \in \mathcal{O}_{\mathrm{reg}}$ and in that case there always exists a constant $g \in G_{\mathbb{C}}$ such that $g E g^{-1} = E_{[N]}$ ($= \sum_{i=1}^{N-1} E_i^+$).

\subsection{Monopole Solutions}
To include the presence of knots, we additionally want to add a mo\-no\-pole-\-like behaviour near the points $p_a$ at which $K$ intersects $\Sigma$.
Monopoles are characterized by the fact that they exhibit a monodromy of magnetic charge $\lambda \in \Gamma_{\mathrm{char}}^\vee$ around the origin in $\mathbb{C}$.
The monodromy is carried by the behaviour of $\varphi$ when moving in a circle around the origin $z=r e^{i\vartheta} \mapsto r e^{i(\vartheta+2\pi)}$.
This is encoded in the following knot singularity ansatz
\begin{align}
	A^\lambda =0, \qquad 	
	\varphi^\lambda =\sum_{i=1}^{N-1} z^{\lambda_i}  E_i^+, \qquad
	\phi_1^\lambda = 0.
	\label{eq:khovanov-knot-singularity-ansatz}
\end{align}
This ansatz provides an initial solution of the $\mathcal{G}_{\mathbb{C}}$-invariant part of the EBE $[\mathcal{D}_i, \mathcal{D}_j] = 0$, $i=1,2,3$.

In this ansatz $\varphi^\lambda$ is an element of $\mathcal{O}_{\mathrm{reg}}$ for all $z\neq 0$, but exactly at $z=0$ it is an element of some subordinate orbit $\mathcal{O}_\pi$.
The partition $\pi$ is given by the Jordan blocks of $\varphi^\lambda{}\restr_{z=0}$.
It can be determined from the weight $\lambda$ by moving through the entries of $\lambda = (\lambda_1, \ldots, \lambda_{N-1})$, counting the number of consecutive $\lambda_i$ with value 0, and shifting the resulting counts by +1.
For example, if the knot is labeled by a co-character that corresponds to either the fundamental or anti-fundamental representation, one finds:
\begin{align}
	\lambda = (\underbracket{\hspace{.5em}\strut}_{0}\hspace{-.2em} 1, \hspace{.2em} \underbracket{\, 0, \ldots, 0\ \strut}_{N-2} ) 
	\quad  \text{or} \quad 
	(\, \underbracket{0,\ldots, 0, \strut}_{N-2} \hspace{.2em} 1 \hspace{-.2em}\underbracket{\hspace{.5em}\strut}_{0} )
	\quad \leadsto \quad 
	\pi = [(N-1)\; 1].
\end{align}
In that case $\varphi^\lambda{}\restr_{z=0}$ is an element of the (unique) subregular nilpotent orbit $\mathcal{O}_{\mathrm{subreg}} = \mathcal{O}_{[N-1\ 1]}$.
Note that the same is true when the knot is labeled by any symmetric or anti-symmetric representation of $\mathfrak{sl}(N,\mathbb{C})$, since these correspond to the co-characters $\lambda = (n,0,\ldots, 0)$ and $(0,0,\ldots, n)$.
Higher (anti-)symmetric representations are distinguishable from the fundamental ones because the associated monodromies have higher winding number.

Let us note for later that the ansatz in~\eqref{eq:khovanov-knot-singularity-ansatz} is of the form $\varphi^\lambda = E + K(z,\lambda)$ for some basepoint $E \in \mathcal{O}_{\pi}$ together with a choice of $\mathfrak{sl}_2$-completion $(E,H,F)$, and where $K(z,\lambda) \in \ker F \cap \mathcal{N}$ is a holomorphic function that vanishes at $z=0$.
Put differently, $\varphi^\lambda$ is a map from $\mathbb{C}$ to the Slodowy slice $S_E \cap \mathcal{N}$ that sends $z=0$ to the basepoint $E$ and monodromy prescribed by $\lambda$.

Consider now a complex gauge transformation $g_\lambda \in \mathcal{G}_{\mathbb{C}}(\mathbb{C}\times\mathbb{R}^+_y)$ and assume it is of the form $g_\lambda = \exp \psi$ for some $\psi \in i\mathfrak{h}$.
The moment map condition becomes
\begin{align}
	(\Delta_{z,\bar z} + \partial_y^2)\ \psi + \frac{1}{2} [\overline{\varphi}^\lambda , e^{2\psi} \varphi^\lambda e^{-2\psi}] = 0.
	\label{eq:khovanov-moment-map-after-gauge-trafo-3d}
\end{align}
Here $\Delta_{z,\bar z}= 4 \partial_z \partial_{\bar z}$ denotes the Laplacian on $\mathbb{C}$.
Note that this is simply the three-dimensional version of~\eqref{eq:khovanov-moment-map-after-gauge-trafo-1d}.
Since any solution of~\eqref{eq:khovanov-moment-map-after-gauge-trafo-3d} gives rise to a solution of the extended Bogomolny equations, we will abbreviate this equation by $\EBE(\psi) = 0$.

Mikhaylov proved that, for $G=SU(N)$ and any weight $\lambda\in\Gamma_{\mathrm{char}}^\vee$, there exists a unique solution $g_\lambda$ that is compatible with the Nahm pole solution at boundary points away from $z=0$.
The explicit formulae are unfortunately somewhat unwieldy; we refer to \cite{Mikhaylov2012} for a detailed description of the general case and only state the results for $G=SU(2)$.

For $G=SU(2)$, the Cartan subalgebra is spanned by a single element $H$ and $\lambda$ is a single non-negative half-integer.
In that case $g_\lambda$ is comparatively simple.
In spherical coordinates $(R, \psi, \vartheta)$ on $\mathbb{C} \times \mathbb{R}^+_y$ and using $s = \sin(\pi/2-\psi)$ as boundary defining function on $\mathbb{C}\setminus\{0\} \times \mathbb{R}^+_y$, $g_\lambda$ is given by
\begin{align}
	g_\lambda &= \exp\bigg(\ \log \frac{\lambda+1}{R^{\lambda+1} s s_\lambda}\; H\ \bigg).
	\label{eq:khovanov-knot-singularity-transformation}
\end{align}
Here we used the abbreviation $s_\lambda = \frac{(1+s)^{\lambda+1} - (1-s)^{\lambda+1}}{2s}$.
The action of $g_\lambda$ on $\mathcal{D}_\mu$ yields the field configuration:
\begin{align}\label{eq:adiabatics-Nahm-Pole-knot-singularity-model-solutions}
\begin{split}
	A_\vartheta &= - (\lambda+1) \cos^2 \psi \frac{(1+\cos \psi)^{\lambda} - (1-\cos\psi)^{\lambda}}{(1+\cos\psi)^{\lambda+1}-(1-\cos\psi)^{\lambda+1}}\ H, \\[1ex]
	\phi_1 &= -\frac{\lambda+1}{R} \frac{(1+\cos\psi)^{\lambda+1} + (1-\cos\psi)^{\lambda+1}}{(1+\cos\psi)^{\lambda+1} - (1-\cos\psi)^{\lambda+1}}\ H, \\[1ex]
	\varphi &= \frac{\lambda+1}{R} \frac{\sin^\lambda \psi\ \exp(i\lambda\vartheta) }{(1+\cos\psi)^{\lambda+1} - (1-\cos\psi)^{\lambda+1}}\ E, \\[1ex]
	A_s &= A_t = A_R = A_\psi = 0.
\end{split}
\end{align}

\subsection{The Nahm Pole Boundary Conditions}
We are now ready to provide a definition of the Nahm pole boundary conditions that is convenient for the discussions in this article.
\begin{definition}[Nahm Pole Boundary Condition]
Consider three-manifolds of the form $\Sigma\times \mathbb{R}^+_y$.
Let $g_0$ and $g_\lambda$ be the singular gauge transformations given in equations~\eqref{eq:khovanov-nahm-pole-singularity-transformation}~and~\eqref{eq:khovanov-knot-singularity-transformation}, respectively.
The fields $(A,\varphi,\phi_1)$ satisfy regular Nahm pole boundary conditions with knot singularities at a divisor $D = \{(p_a, \lambda_a)\}_{a=1,\ldots, k} \subset \Sigma$ if
\begin{itemize}
	\item in a neighbourhood of a boundary point $(p,0) \in (\Sigma\setminus D) \times \mathbb{R}^+_y$ away from knot insertions, there exists a $G_{\mathbb{C}}$-valued gauge transformation of the form $g=g_0 e^u$ with $\abs{u}+\abs{y du} < C y^\epsilon$, such that
	\begin{align}
		(A, \varphi, \phi_1) = g \cdot (0, \sum E_i^+, 0),
	\end{align}
	\item in a neighbourhood of a knot insertion $(p_a, 0) \in \Sigma \times \mathbb{R}^+_y$ of weight $\lambda$, there exists a gauge transformation of the form $g=g_\lambda e^u$ with $\abs{u} + \abs{Rs\; du} \leq C R^\epsilon s^\epsilon$, such that 
	\begin{align}
		(A, \varphi, \phi_1) = g \cdot (0, \sum z^{\lambda_i} E_i^+, 0).
	\end{align}
\end{itemize}
\end{definition}
Up to an inconsequential reinterpretation of field components (cf. \cite{Bleher2024b}), this definition lifts to Nahm pole boundary conditions for Kapustin-Witten fields $(A,\phi)$ on four-manifolds $X^3 \times \mathbb{R}^+_y$ and Haydys-Witten fields $(A,B)$ on five-manifolds $W^4\times \mathbb{R}^+_y$.
In higher dimensions, the knot singularities are supported along a knot $K \subset X^3 \times \{0\}$ or a surface $\Sigma_K \subset W^4 \times \{0\}$, respectively.


\section{EBE-Solutions and Higgs bundles}
\label{sec:khovanov-kobayashi-hitchin-correspondence}

The Hermitian Yang-Mills structure of the extended Bogomolny equations on $\Sigma\times\mathbb{R}^+_y$ suggests that there is a deep relation between full solutions and the simpler holomorphic data that underlies the initial knot singularity ansatz~\eqref{eq:khovanov-knot-singularity-ansatz}.
Indeed, there is a Kobayashi-Hitchin correspondence between solutions of the extended Bogmolny equations on $\Sigma \times \mathbb{R}^+_y$ and Higgs bundle data on $\Sigma$.
This correspondence was originally proposed by Gaiotto and Witten in \cite{Gaiotto2012a} and has since been proven by He and Mazzeo \cite{He2019c, He2020b} (also see \cite{He2019b, Dimakis2022a, Sun2023} for variations of this correspondence).
Here we repeat the relevant definitions, review parts of the proof that will be of particular relevance to us, and use the opportunity to fix some notation.

We are interested in solutions of the EBE on $\Sigma\times\mathbb{R}^+_y$ that, on the one hand, satisfy Nahm pole boundary conditions with knot singularities at the boundary, and for which, on the other hand, $A+i\varphi$ approaches a flat $G_{\mathbb{C}}$ connection as $y\to\infty$.
Let $D = \{ (p_a, \lambda_a) \}_{a=1,\ldots,k}$ denote a collection of points $p_a \in \Sigma$ that are decorated with weights $\lambda_a \in \Gamma_{\mathrm{char}}^\vee$ in the co-character lattice of $\mathfrak{g}$.
The moduli space of EBE-solutions with knot singularity data $D$ will be denoted by:
\begin{definition}[Moduli Space of EBE Solutions]
\begin{align}
	\begin{multlined}
    \widehat{\mathcal{M}}^{\mathrm{EBE}}_{D} := 
		\Big\{\
		\EBE(A,\varphi,\phi_1) = 0 \mathrel{\Big|}
		(A, \varphi, \phi_1) \text{ satisfies Nahm-pole} \\ \text{boundary conditions as } y\to 0,
		\text{has knot singularities at } D, \text{ and }\\
		\text{ approaches an irreducible flat } SL(N, \mathbb{C}) \text{ connection as } y \to \infty
		\Big\}.
		\end{multlined}
\end{align}
\end{definition}
The absence of a knot will be denoted by $D=\emptyset$, in which case $\mathcal{M}^{\mathrm{EBE}}_{\emptyset}$ is the moduli space of pure Nahm pole solutions.

Let $\mathcal{G}_0(\Sigma \times \mathbb{R}^+_y)$ be the subset of real gauge transformations that vanish at the boundary.
Throughout we will denote the quotients of moduli spaces by the action of this gauge group by the same symbols, but without the hat.
For example for the moduli space of EBE solutions with knot singularities at $D$ we have:
\begin{align}
\mathcal{M}^{\mathrm{EBE}}_D = \widehat{\mathcal{M}}^{\mathrm{EBE}}_D \big/ \mathcal{G}_0(\Sigma\times\mathbb{R}^+_y).
\end{align} 
\begin{remark}
We only mod out $\mathcal{G}_0(\Sigma\times\mathbb{R}^+_y)$ because the configuration of the fields at the boundary is in principle a physical observable; relatedly, from the perspective of mathematics, the boundary configuration is part of the input of the variational principle.
\end{remark}

We have seen in the previous section, for $\Sigma = \mathbb{C}$, that once the data of an initial solution of $[\mathcal{D}_i, \mathcal{D}_j] = 0$ is fixed, the remaining real moment map equation determines a unique complex gauge transformation, $g_0$ or $g_\lambda$, that transforms the ansatz into a proper solution of the extended Bogomolny equations.
The Kobayashi-Hitchin correspondence states that this generalizes to arbitrary Riemann surfaces $\Sigma$, where the initial solution is replaced by certain holomorphic data over $\Sigma$.
In the following, we introduce the relevant geometric objects over Riemann surfaces $\Sigma$.

Let $(\mathcal{E}, \bar\partial_{\mathcal{E}})$ be a holomorphic vector bundle over $\Sigma$.
We write $\Omega^{1,0}(\End \mathcal{E})$ for the space of holomorphic one-forms with values in the endomorphism bundle of~$\mathcal{E}$.
\begin{definition}[Higgs Bundle]
A Higgs bundle $(\mathcal{E}, \varphi)$ is a holomorphic vector bundle $(\mathcal{E}, \bar\partial_{\mathcal{E}})$ over $\Sigma$, together with a holomorphic one-form $\varphi\in \Omega^{1,0}(\End \mathcal{E})$ called the Higgs field.
\end{definition}
We will always assume that $\det \mathcal{E} = \mathcal{O}_\Sigma$, the sheaf of holomorphic functions on $\Sigma$, and that $\deg \mathcal{E} = 0$.
In that situation, a Higgs bundle is called stable if $\deg V < 0$ for any holomorphic subbundle $V$ of $\mathcal{E}$ that satisfies $\varphi(V) \subset V\otimes K$ and polystable if it is a direct sum of stable Higgs bundles.

We denote the moduli space of Higgs bundles by $\widehat{\mathcal{M}}_{\mathrm{Higgs}}$ and as above write $\mathcal{M}_{\mathrm{Higgs}}$ for its quotient by $SL(N,\mathbb{C})$-valued gauge transformations $\mathcal{G}_{\mathbb{C}}(\Sigma)$.
By a famous result of Hitchin, for any Higgs bundle there exists an irreducible solution of the Hitchin equations if and only if it is stable (and a reducible solution if and only if polystable) \cite{Hitchin1987a}.
Instead of introducing the Hitchin equations, let us simply state that a solution of the Hitchin equations is equivalent to a flat $SL(N, \mathbb{C})$ connection, which in turn are classified by $\rho:\pi_1(\Sigma) \to SL(N,\mathbb{C})$.
A series of articles by Hitchin, Donaldson, Simpson, and Corlette famously culminated in a proof that there is a diffeomorphic equivalence between the moduli spaces of stable Higgs bundles, irreducible solutions of the Hitchin equations, and irreducible flat connections \cite{Donaldson1987a, Corlette1988, Hitchin1987, Simpson1990, Simpson1992}.

In the study of Higgs bundles an important role is played by the Hitchin fibration.
It is built from the adjoint quotient map, which sends a matrix in $\mathfrak{sl}(N,\mathbb{C})$ to its generalized eigenvalues:
\begin{align}
	\chi : \mathfrak{sl}(N, \mathbb{C}) \to \mathfrak{h}/\mathcal{W}, \
	A \mapsto (c_2(A),\ldots, c_{N}(A) ),
\end{align}
where $\mathfrak{h}$ is the Cartan subalgebra, $\mathcal{W}$ the Weyl group, and $c_j(A)$ denote the invariant polynomials of $\mathfrak{sl}(N,\mathbb{C})$ as determined by $\det(\lambda1 - A) = \sum \lambda^{N-j}(-1)^j c_j(A)$.
The Hitchin fibration lifts this to the level of Higgs bundles~\cite{Hitchin1987a}.
\begin{align}
	\mathcal{M}_{\mathrm{Higgs}} &\to \bigoplus_{i=1}^n H^0( \Sigma, K^{i+1} ), \\
	(\mathcal{E},\varphi) &\mapsto (c_2(\varphi),\ldots, c_{N}(\varphi) ).
\end{align}
The Hitchin component (or Hitchin section) $\mathcal{M}_{\mathrm{Hit}}$ is the section of this fibration that associates to the invariants $(q_2, \ldots q_{N})$ the gauge orbit of the following Higgs bundle
\begin{align}
	\mathcal{E} = K^{-\frac{N-1}{2}} \oplus K^{-\frac{N-1}{2} +1} \oplus \ldots \oplus K^{\frac{N-1}{2}}, \qquad
	\varphi = \begin{pmatrix} 0 & * & 0 & \\   & 0 & \ddots & 0  \\ && \ddots &  * \\ q_{N} & \ldots & q_2 & 0\end{pmatrix}.
\end{align}

Given a Higgs bundle $(\mathcal{E}, \varphi)$ and a holomorphic line subbundle $L \subset \mathcal{E}$, there is an associated divisor $\mathfrak{d} = \mathfrak{d}(L,\varphi)$ that encodes the linear dependencies between the subbundles $L$, $\varphi(L)$, \ldots, $\varphi^n(L)$ of $\mathcal{E}$.
To make this precise, introduce the following maps
\begin{align}
	f_i := 1\wedge \varphi \wedge \ldots \wedge \varphi^i &: L^i \to K^{\frac{i(i+1)}{2}} \otimes L^{-(i+1)} \otimes \wedge^{i+1} \mathcal{E}.
\end{align}
Note that these maps capture the linear dependencies between the subbundles $L, \varphi(L), \ldots \varphi^n(L)$ through their zeroes, by virtue of being antisymmetric.
The divisor associated to $(L,\varphi)$ is constructed from the zeroes of $f_i$ and their vanishing orders as follows:
\begin{align}
	\mathfrak{d}(L,\varphi) &:= \sum_{\substack{i=1,\ldots, n \\ p \in \Sigma}} \operatorname{ord}_{p} f_i\ \cdot p.
\end{align}
Following \cite{He2019c, He2020b}, we call $\mathfrak{d}(L,\varphi)$ effective if at each point $p$ the tuple
\begin{align}
	\lambda = \big( \operatorname{ord}_p f_1\; ,\ \operatorname{ord}_p f_1 - \operatorname{ord}_p f_2\; ,\ \ldots,\ \operatorname{ord}_p f_{n-1} - \operatorname{ord}_p f_n \big)
\end{align}
has only non-negative entries.
In that case we also denote the divisor by $\mathfrak{d}(L,\varphi) = \{ (p_a, \lambda_a) \}_{a=1,\ldots, k}$.
\begin{definition}[Effective Triples]
An effective triple $(\mathcal{E}, \varphi, L)$ consists of a stable Higgs bundle $(\mathcal{E}, \varphi)$ together with a holomorphic line bundle $L$ such that the divisor $\mathfrak{d}(L,\varphi)$ is effective.
\end{definition}
The moduli space of effective triples will be denoted $\widehat{\mathcal{M}}_{(\mathcal{E},\varphi,L)}$ and we also write $\mathcal{M}_{(\mathcal{E},\varphi,L)}$ for its quotient by $\mathcal{G}_{\mathbb{C}}(\Sigma)$.

We have now collected all ingredients to state the Kobayashi-Hitchin correspondence for solutions of the extended Bogomolny equations on $\Sigma\times\mathbb{R}^+_y$ with Nahm pole boundary conditions (NP) and Nahm pole boundary conditions with knots (NPK).
\begin{theorem}[\cite{He2019c, He2020b}]
\label{thm:khovanov-Gaiotto-Witten-bijections}
There are bijections
\begin{align}
	\mathcal{M}^{\mathrm{EBE}}_{\emptyset} \xrightarrow{I_{NP}} \mathcal{M}_{\mathrm{Hit}},
	\qquad
	\mathcal{M}^{\mathrm{EBE}}_{D} \xrightarrow{I_{NPK}} \mathcal{M}_{(\mathcal{E},\varphi, L)}.
\end{align}
\end{theorem}
For recent variants of this result on $\Sigma = \mathbb{C}$ we also refer to \cite{Dimakis2022a}, who elaborated on the situation for nilpotent Higgs fields, as well as \cite{Sun2023}, for the case that $\phi_1$ approaches a non-zero value at $y\to \infty$, a situation that is also known as ``real symmetry breaking'' \cite{Gaiotto2012a}.
Since the construction of $I_{NP}$ and $I_{NPK}$ will play an important role in the remainder of this section, we will now provide a brief review.

As a start, one observes that any Nahm pole solution of the EBE equations determines an effective triple.
This was originally explained by Gaiotto and Witten and can be seen as follows.
For the moment, write $V$ for the vector bundle over $\Sigma\times\mathbb{R}^+_y$ associated to the $N$-dimensional fundamental representation of $G_{\mathbb{C}} = SL(N,\mathbb{C})$.
Denote by $V_y$ the restriction of $V$ to $\Sigma \times \{y\}$.
Observe that $\mathcal{D}_1$ provides a $\bar\partial$-operator on $V_y$ that satisfies $\bar\partial^2 = 0$.
By the Newlander-Nirenberg theorem, this makes $V_y$ into a holomorphic vector bundle that we will denote by $\mathcal{E}_y$.
Next, $\mathcal{D}_3 = \ad_\varphi$ can be interpreted as a $K_\Sigma$-valued endomorphism of $\mathcal{E}_y$ and, moreover, $[\mathcal{D}_1, \mathcal{D}_3] = 0$ implies that $\mathcal{D}_1 \varphi = 0$, so this endomorphism is holomorphic.
Put differently, if we let $\varphi_y$ be the restriction of $\varphi$ to $(\ad E_{\mathbb{C}})_y$, we obtain a family of Higgs bundles $(\mathcal{E}_y, \varphi_y)$ over $\Sigma$.
Finally, $\mathcal{D}_2 = \nabla^A_y - i [\phi_1, \cdot]$ provides a notion of parallel transport in the $y$-direction of $\Sigma\times \mathbb{R}^+_y$.
The equations $[\mathcal{D}_1,\mathcal{D}_2] = [\mathcal{D}_2, \mathcal{D}_3] = 0$ then imply that the family of Higgs bundles is parallel with respect to $\mathcal{D}_2$.

The boundary conditions at $y\to 0$ and $y\to\infty$ provide two additional points of data as follows.
On the one hand, the asymptotic boundary condition at $y\to\infty$, namely that $(A, \varphi, \phi_1)$ converges to an irreducible flat $SL(N,\mathbb{C})$ connection, is equivalent to the statement that the one-parameter family $(\mathcal{E}_y, \varphi_y)$ consists of stable Higgs bundles.
Since the one-parameter family is parallel with respect to $\mathcal{D}_2$, it is then fully determined by specifying the limiting stable Higgs bundle $(\mathcal{E}_\infty, \varphi_\infty)$.

On the other hand, the Nahm pole boundary condition at $y=0$ with knot singularity data $D = \{ (p_a, \lambda_a) \}_{a=1,\ldots, k}$ determines a distinguished line bundle $L \to \Sigma \times \mathbb{R}^+_y$ as follows.
Let $\{U_a\}_{a=0,\ldots k}$ be a collection of open disks, with $U_0$ an open set that does not contain any $p_a$ and the remaining $U_a$ centered at $p_a$, respectively, and such that $\bigcup_{a=0}^k U_a = \Sigma$.
Use spherical coordinates $(R_a, \vartheta_a, \psi_a)$ with boundary defining function $s=\sin(\pi/2 - \psi)$ on $U_a \times \mathbb{R}^+$.
\begin{align}
	L :=
		\begin{multlined}[t]
		\big\{
			u \in \Gamma(\Sigma\times\mathbb{R}^+_y, \ad E_{\mathbb{C}} )\ \mathrel{\big|} \
			\mathcal{D}_2 u = 0\; ,\ \lim_{y\to 0} | y^{-(N-1)/2+\epsilon} u | = 0\ \text{on } U_0, \\
			\text{ and } \lim_{s \to 0} | \psi_a^{-(N-1)/2+\epsilon} u | = 0\ \text{on } U_a,\ a=1,\ldots,k,\ \epsilon>0
		\big\}.
		\end{multlined}
\end{align}
This is a line bundle by definition of the Nahm pole boundary condition.
To see this note that $\mathcal{D}_2 u = \partial_y u - i\phi_1 u = 0$.
Away from knot insertions $\phi_1 = H/y$ has eigenvalues $(N-1)/2y$, $(N-2)/2y$, \ldots, $-(N-1)/2y$, such that there is only a single component of $u$ whose parallel transport vanishes at the maximal possible rate $y^{(N-1)/2}$.
At a knot insertion $p_a$, the same argument with the version of $\phi_1$ for a monopole solution shows that the maximal vanishing rate is $\psi_a^{-(N-1)/2}$.
$L$ is commonly called the \emph{vanishing line bundle}.
For each $y\in \mathbb{R}^+_y$ it is a holomorphic subbundle of $\mathcal{E}_y$.

In summary, the vanishing line bundle over $\Sigma\times\mathbb{R}^+$ determines a unique holomorphic line subbundle of $\mathcal{E}$.
Moreover, the divisor $\mathfrak{d}(L,\varphi)$ is effective and at each point $p$ the zeroes of $f_i$ are exactly such that their vanishing orders are related to the co-character by $\lambda = ( \operatorname{ord}_p(f_1), \operatorname{ord}_p(f_2) - \operatorname{ord}_p(f_1), \ldots )$.
In absence of knot singularities the divisor is empty and one can show that the Higgs bundle is an element of the Hitchin section.

Proving that, conversely, each effective triple $(\mathcal{E}, \varphi, L)$ gives rise to a solution of the extended Bogomolny equations, involves more work.
This part of the theorem is due to He and Mazzeo and we refer to \cite[Sec. 7]{He2020b} for more details.
Since we later find that similar arguments should apply to solutions of the decoupled Haydys-Witten equations, we summarize the basic approach while omitting the necessary analytic prerequisites.

Assume we are given an effective triple $(\mathcal{E}, \varphi, L)$.
From the discussion above it's clear that the effective divisor $\mathfrak{d}(L,\varphi)$ determines the position and charges $\{(p_a, \lambda_a)\}_{a=1,\ldots, k}$ of knot singularities.
The main insight is that the effective triple provides enough information to construct a field configuration $(A, \varphi, \phi_1)$ on $\Sigma\times\mathbb{R}^+_y$ that satisfies the Nahm pole boundary conditions with knot singularities and is a solution of the EBE at leading order in $y^{-1}$.
This approximate solution can subsequently be improved, order by order in $y$, to a unique, proper solution of the EBE.
For the construction we choose an open cover of $\Sigma$, consisting of an open set $U_0$ that does not contain any of the points $p_a$ and non-intersecting open disks $U_a$ centered at $p_a$.
This is used to construct field configurations on each of the $U_j$ independently, which are then glued over $U_0$ to an approximate solution on $\Sigma$.
The construction proceeds in four steps.

First, dropping the index $a$ for the moment, we restrict to a small disk $U$ centered at $p$.
We need to extract from $(\mathcal{E},\varphi,L)$ a field configuration on $U \times \mathbb{R}^+_y$ that looks like the knot singularity ansatz $(0,\varphi = \sum_i z^{\lambda_a} E_i^+, 0)$.
Let $z$ be a holomorphic coordinate on $U$ and use spherical coordinates $(R,\psi,\vartheta)$ on $U \times \mathbb{R}^+_y$.
Write $\varphi = \varphi_z dz$ and set $L_1 := L$ and $L_{i+1} := \varphi_z (L_i)$.
Choose a non-vanishing section $e_1$ of $L$ and extend it to a local holomorphic frame $\{e_1, \hat{e}_2, \ldots \hat{e}_{n+1}\}$.
Let $\lambda_1$ be the order of vanishing of $1 \wedge \varphi_z: L^2 \to \wedge^2 \mathcal{E}$.
We can now write $\varphi_z(e_1) = f e_1 +  z^{\lambda_1} \sum_{i=2}^{n+1} c_i \hat{e}_i$, where at least one of the $c_i$ is non-vanishing at $z=0$, because the map $1\wedge \varphi_z \wedge \ldots \wedge \varphi_z^n : L_1 \wedge L_2 \wedge \ldots L_{n+1} \to \det \mathcal{E}$ fails to be an isomorphism exactly at $z=0$.
Setting $e_2 := \sum_{i=2}^{n+1} c_i \hat{e}_i$, we have arranged that $\varphi_z(e_1) = f e_1 + z^{\lambda_1} e_2$.
Next, let $\lambda_2$ be the order of vanishing of $1\wedge\varphi_z\wedge\varphi_z^2$, such that $\varphi_z(e_2) = f_1 e_1 + f_2 e_2 + z^{\lambda_2} \sum_{i=3}^{n+1} c_i \hat{e}_i$, and proceed by induction.
We obtain a frame $\{e_1, \ldots, e_{n+1}\}$ for which $\varphi_z(e_i) = z^{\lambda_a} e_{i+1} + \spans{e_1, \ldots e_i}$
Equivalently, when viewed as an $\End{\mathcal{E}}$-valued function, $\varphi_z = \sum z^{\lambda_a} E_i^+$, as required.

In the second step, we act on $(0, \sum z^{\lambda_i} E_i^+ ,0)$ with the singular gauge transformation $g_\lambda$ of \autoref{eq:khovanov-knot-singularity-transformation}.
The result is of order $\mathcal{O}(R^{-1}s^{-1})$, as is required for a field configuration that satisfies Nahm pole boundary conditions with knot singularity at $p$ of weight $\lambda$.
Note, in particular, that when $R\neq0$ and $s\to0$, the gauge transformation is of order $(Rs)^{-1} = y^{-1}$ and is asymptotically equivalent to $g_0$.

The third step constructs a global gauge transformation $g$ over $\Sigma\times\mathbb{R}^+_y$ from the collection of local frames and gauge transformations $\{ g_{\lambda_a} \}$ for each $U_a\times\mathbb{R}^+_y$ .
For this, choose a holomorphic frame on $U_0$ and denote by $g_0$ the gauge transformation defined in equation~\eqref{eq:khovanov-nahm-pole-singularity-transformation}.
On the overlaps $U_0 \cap U_a$, the holomorphic frames are related by an explicit transition function, given by
\begin{align}
	g_{0a} = \exp\big( - \log r \sum_{i=1}^n  \lambda_{a,j} A_{ij}^{-1}  H_i \big).
\end{align}
We find that in the limit $s_a\to 0$ each $g_{\lambda_a}$ is equivalent to $g_0 e^u$ with $\abs{u}+\abs{y du} < C y^\epsilon$.
Gluing the various gauge transformations with help of a partition of unity produces a global gauge transformation $g$ over $\Sigma$.
By construction, $g$ determines a field configuration that satisfies Nahm pole boundary conditions with knot singularities along $K$.

In the final step, the approximate solution is improved to a proper solution of the equations.
For this, one first improves the approximation near knot singularities for each $U_a\times\mathbb{R}^+_y$ by solving the moment map equation to desired precision, order by order in $R$.
Determining in that way the higher order terms of $g_{\lambda_a}$, and if necessary using Borel resummation to find a function that has the given expansion, one obtains a gauge transformation $g$ on all of $\Sigma\times\mathbb{R}^+_y$ for which the solution near $p_a$ vanishes to all orders as $R_a\to 0$.
Carrying out an analogous procedure for the higher orders of $g$ in an expansion in $y$ yields a unique solution of the extended Bogomolny equations.


\section{The Isotopy Ansatz}
\label{sec:khovanov-isotopy-ansatz}

In this section we return to investigate the decoupled Haydys-Witten equations over $M^5 = C \times \Sigma \times \mathbb{R}^+_y$.
In contrast to the analysis of the extended Bogomolny equations reviewed in the preceding sections, we now investigate situations in which $\mathcal{D}_0$ is non-zero.
For the time being, we assume that $C = \mathbb{R}_s \times S^1_t$ and restrict ourselves to the investigation of $\mathbb{R}_s$-invariant solutions of the decoupled Haydys-Witten equations in temporal gauge $A_s = 0$.
This corresponds to a dimensional reduction of the decoupled Haydys-Witten equations for which $\mathcal{D}_0 = i D_t$.
The result are differential equations over the four-manifold $S^1_t \times \Sigma \times \mathbb{R}^+_y$ that correspond to a decoupled version of the Kapustin-Witten equations.

We start here with an analogue of the considerations in \autoref{sec:khovanov-Nahm-pole-boundary-condition}, where we described the Nahm pole and monopole model solutions.
Let $\Sigma = \mathbb{C}$ with complex coordinate $z = x^2+ix^3$.
Consider a single-stranded knot $K$ that extends along $S^1_t$, and view it as the image of
\begin{align}
	\beta: S^1_t &\to S^1_t \times \mathbb{C} \times \mathbb{R}^+, \\
	t &\mapsto (t, z_0(t),0).
\end{align}
There always exists an isotopy $\beta_\bullet$ that connects $\beta_0(t) = (t,0,0)$, the $S^1$-invariant knot centered at the origin in $\mathbb{C}$, to $\beta_1(t) = \beta(t)$.
To be specific, let us choose the isotopy
\begin{align}
	\beta_\bullet : & [0,1]_q \times S^1_t \rightarrow S^1_t \times \mathbb{C} \times \mathbb{R}^+_y, \\
	& (q,t) \mapsto (t, qz_0(t), 0).
\end{align}
For fixed $q\in[0,1]$, introduce the comoving holomorphic coordinate $\zeta_q(t)=z-qz_0(t) \in \mathbb{C}$.
The knot defined by $\beta_q$ lies at the origin $\{\zeta_q=0\}$ of the complex plane parametrized by $\zeta_q$.
Let us also introduce polar coordinates $\zeta_q = r_q e^{i\vartheta_q}$.
We often drop the subscript $q$ from $\zeta_q$, $r_q$ and $\vartheta_q$ and only highlight the dependence on $q\in[0,1]$ where necessary.

\begin{figure}
\centering
\begin{minipage}{5cm}
\includegraphics[width=5cm, height=5cm, keepaspectratio]{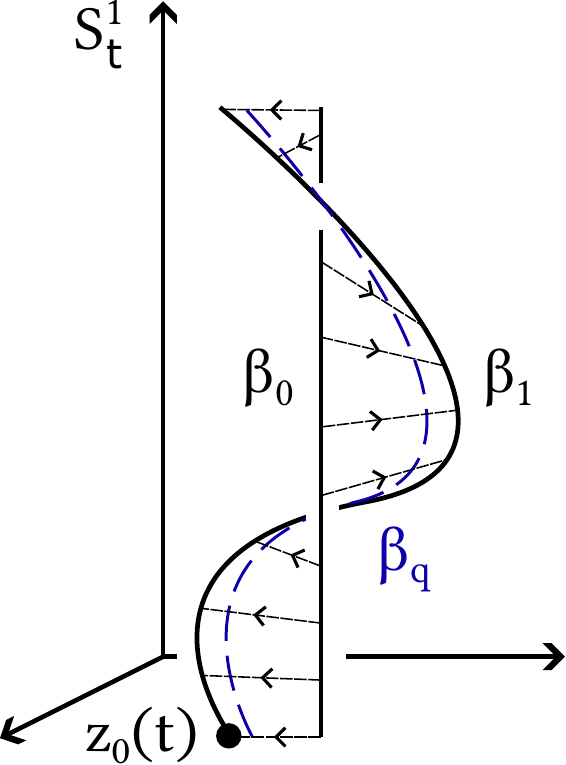}
\end{minipage}
\begin{minipage}{4cm}
\includegraphics[width=4cm, height=4cm, keepaspectratio]{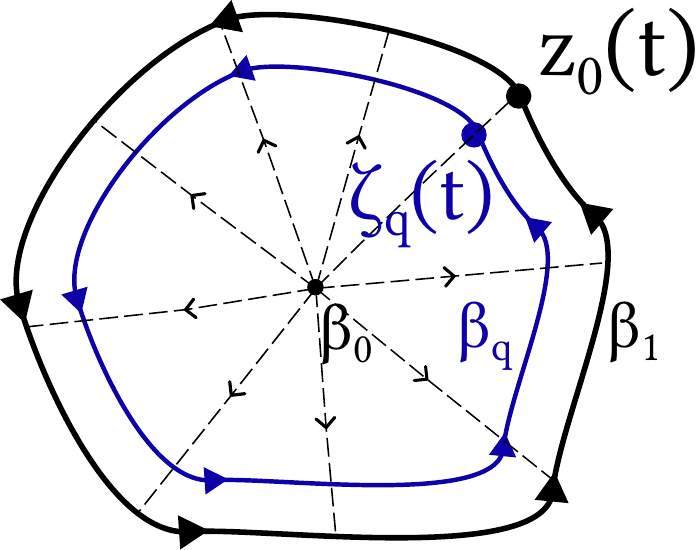}
\end{minipage}
\caption{Isotopy $\beta_\bullet$ interpolating between the $S^1_t$-invariant strand $\beta_0$ centered at the origin of $\mathbb{C}$ and a non-trivial single stranded knot $\beta_1$. 
The isotopy describes a homotopy of trajectories $\zeta_q(t)$ that interpolate from the constant to the original trajectory $z_0(t)$.}
\end{figure}

Our starting point is the following generalization of the ansatz $(A^\lambda, \varphi^\lambda, \phi_1^\lambda)$ in equation~\eqref{eq:khovanov-knot-singularity-ansatz} that was the starting point in \autoref{sec:khovanov-Nahm-pole-boundary-condition} to find the monopole model solutions:
\begin{align}
	A^\beta_t &= \frac{q \dot z_0}{\zeta} \sum_{i,j}^{N-1} \lambda_i A^{-1}_{ij} H_j,  &
	A^\beta_z &= A^\beta_y = 0, \\
	\varphi^\beta &= \sum_{i=1}^{N-1} \zeta^{\lambda_i}  E_i^+, &
	\phi_1^\beta &= 0.
	\label{eq:khovanov-isotopy-ansatz}
\end{align}
Here $\dot z_0$ denotes the derivative of $z_0(t)$ with respect to $t$, which is fully determined by the original knot trajectory and part of the fixed boundary conditions.
For each $q\in[0,1]$, this model solution provides an initial solution of the $G_{\mathbb{C}}$-invariant part of the decoupled Haydys-Witten equations ${[\mathcal{D}_\mu, \mathcal{D}_\nu]=0}$.
We call $(A^\beta, \varphi^\beta, \phi_1^\beta)$ the isotopy ansatz for the decoupled Kapustin-Witten equations.

The singular behaviour of $\varphi^\beta$ at $\zeta=0$ exactly encodes the presence of a single stranded 't Hooft operator of charge $\lambda \in \Gamma_\mathrm{char}^\vee$ along $\beta_q=(t,qz_0(t))$.
In particular, for any fixed $t \in S^1_t$, the terms at order $q^0$ are equivalent to the initial ansatz $(A^\lambda, \varphi^\lambda, \phi^\lambda)$ of~\eqref{eq:khovanov-knot-singularity-ansatz} with knot singularity at $z=qz_0(t)$.
Corrections due to the $t$-dependence of the knot singularity arise only at order $\mathcal{O}(q^1)$.
Note, in particular, that for $q=0$ the ansatz genuinely coincides with $(0, \sum_i z^{\lambda_i} E_i^+ , 0)=(A^\lambda, \varphi^\lambda, \phi^\lambda)$, where the knot is $S^1$-invariant and located at $\zeta_{q=0} = z =0$.

We propose that, by a continuity argument along the isotopy parameter $q\in[0,1]$, one can find a complex gauge transformation $g_\beta \in\mathcal{G}_{\mathbb{C}}(S^1_t \times \mathbb{C} \times \mathbb{R}^+_y)$ that, on the one hand, solves the moment map condition $\sum [\overline{\mathcal{D}}_\mu, \mathcal{D}_\mu]=0$ and, on the other hand, encodes Nahm pole boundary conditions.

For this, apply a complex gauge transformation $g_\beta = \exp \psi$ with $\psi \in i\mathfrak{h}$ to the differential operators $\mathcal{D}_\mu$ obtained from the isotopy ansatz $(A^\beta,\varphi^\beta,\phi_1^\beta)$.
After this gauge transformation, the moment map condition is equivalent to $\sum [\overline{\mathcal{D}}_\mu, e^{-2\psi} \mathcal{D}_\mu e^{2\psi}] = 0$.
The summand $[\overline{D_0}, e^{-2\psi} D_0 e^{2\psi}]$ reduces to an expression of the form $\ad_{X}$, where most terms in $X$ cancel because $A_t$ and $\psi$ are both elements of the Cartan subalgebra $\mathfrak{h}$:
\begin{align}
  X &=
	\begin{multlined}[t]
		e^{-2\psi} (\partial_t A_t - [A_t, 2 \partial_t \psi]) e^{2\psi} + 2 e^{-2\psi }\partial_t^2 \psi e^{2\psi} \\
		 {}-{} (\partial_t A_t - [A_t, 2 \partial_t \psi]) + [A_t, e^{-2\psi} A_t e^{2\psi}]
	\end{multlined}\\
  &= 2 \partial_t^2 \psi,
\end{align}
where we have used that $e^{2\psi}$ acts trivially on $\mathfrak{h}$.
Furthermore, since $A_z^\beta=A_y^\beta = \phi_1^\beta = 0$, we arrive at
\begin{align}
	\dKW(\psi) =
	(\partial_t^2 + \Delta_{z,\bar z} + \partial_y^2)\ \psi
	+ \tfrac{1}{2} [ \overline{\varphi}^\beta , e^{2\psi} \varphi^\beta e^{-2\psi}]
	= 0.
	\label{eq:khovanov-moment-map-after-gauge-trafo-4d}
\end{align}
Note that~\eqref{eq:khovanov-moment-map-after-gauge-trafo-4d} is just the four-dimensional version of the one-dimensional Nahm equation~\eqref{eq:khovanov-moment-map-after-gauge-trafo-1d} and the three-dimensional extended Bogomolny equation $\EBE(\psi)=0$ in~\eqref{eq:khovanov-moment-map-after-gauge-trafo-3d}.

To keep notation at a minimum we restrict the following discussion to $G=SU(2)$, though the general case with $G=SU(N)$ is not much different.
Assume $\psi = \psi(t,z,y) H$, $\varphi^\beta = \zeta^\lambda E$, and $\overline{\varphi} = \bar{\zeta}{}^\lambda F$, where $(E,H,F)$ is the standard basis of $\mathfrak{sl}(2,\mathbb{C})$.
We can then bring~\eqref{eq:khovanov-moment-map-after-gauge-trafo-4d} into the slightly more explicit form
\begin{align}
	\label{eq:khovanov-moment-map-after-gauge-trafo-4d-v2}
	\dKW[q](\psi) = (\partial_t^2 + \frac{1}{2} \Delta_{z, \bar z} + \partial_y^2) \psi - r_q^{2\lambda} \exp(2\psi) = 0.
\end{align}

Drawing inspiration from Gaiotto-Witten's adiabatic approach, we now restrict to functions $\psi(\zeta,y; z_0(t))$ that depend on $t$ only through $z_0(t)$ and its appearance in the comoving coordinate $\zeta=z-qz_0(t)$.
The operator $\dKW[\bullet]$ describes a homotopy of differential operators and associated boundary conditions.
It interpolates between the operator $\dKW[0] = \EBE$ together with an $S^1$-invariant knot singularity along $K=(t,0,0)$ on the one hand, and the decoupled Kapustin-Witten equations $\dKW[q=1]$ with a knot singularity along $K=(t,z_0(t),0)$ on the other.

Given the assumption that $\psi$ depends adiabatically on $t$, we can replace $\Delta_{z,\bar z} = \Delta_{\zeta, \bar\zeta}$ and split the differentiation with respect to $t$ into its contributions from the comoving coordinate $\zeta$ and explicit appearances of $z_0(t)$:
\begin{align}
	\partial_t^2  = 
			\tilde \partial_t^2 
		- q\ (\ddot z_0 \partial_\zeta + \ddot {\bar z}_0 \partial_{\bar\zeta}) 
		- 2q\ (\dot z_0 \partial_\zeta + \dot{\bar z}_0 \partial_{\bar\zeta})\ \tilde \partial_t 
		+ q^2\ (\dot z_0 \partial_\zeta + \dot {\bar z}_0 \partial_{\bar\zeta})^2.
\end{align}
The notation $\tilde \partial_t$ on the right hand side shall reflect that this derivative only acts on $z_0(t)$ (and its derivatives).

Observe that equation \eqref{eq:khovanov-moment-map-after-gauge-trafo-4d} naturally organizes into powers of $q$.
Accordingly, we make the formal ansatz $\psi=\sum_{n\geq 0} q^n \psi^{(n)}$.
Plugging this into~\eqref{eq:khovanov-moment-map-after-gauge-trafo-4d} and expanding the exponential function in powers of $q$ leads to:
\begin{align}
	\dKW[q](\psi) = \EBE[q](\psi^{(0)}) + \sum_{n \geq 1} q^n\ \dKW_q^{(n)} \big( \psi^{(n)};\ \psi^{(0)}, \ldots, \psi^{(n-1)} \big).
	\label{eq:khovanov-homotopy-expansion-of-dKW}
\end{align}
We call this the homotopy expansion of $\dKW[\bullet]$ induced by the knot isotopy $\beta$
The operator at zeroth order of the expansion is a comoving, or ``adiabatic'', version of the extended Bogomolny equations in equation~\eqref{eq:khovanov-moment-map-after-gauge-trafo-3d}:
\begin{align}
	\EBE[q](\psi^{(0)}) = (\Delta_{\zeta, \bar\zeta} + \partial_y^2 )\ \psi^{(0)} - r_q^{2\lambda} \exp(2\psi^{(0)}).
\end{align}
Corrections to the adiabatic operator appear at order $q^n$ with $n\geq 1$ and are given by the following linear second-order operators
\begin{align}
	\dKW_q^{(n)}&(\psi^{(n)};\ \psi^{(0)}, \ldots, \psi^{(n-1)}) = L_q \psi^{(n)} + K_q^{(n)}(\psi^{(0)}, \ldots, \psi^{(n-1)}).
\end{align}
The linear differential operator $L_q$ is a comoving version of~\eqref{eq:khovanov-moment-map-after-gauge-trafo-4d-v2} and independent of $n$
\begin{align}
	L_q &= \tilde{\partial}_t^2 + \Delta_{\zeta,\bar\zeta} + \partial_y^2 - r^{2\lambda} \exp(2\psi^{(0)}).
\end{align}
In contrast, the inhomogeneous terms $K_q^{(n)}$ depend explicitly (and non-linearly) on the solutions of the lower order equations.
Using the notation $\pi \vdash n$ for a partition of $n$ into $\abs{\pi}=s$ positive integers $\pi_i$ of multiplicity $\nu_i$, the inhomogeneous terms are given by
\begin{align}
	K_q^{(n)} &=
		\begin{multlined}[t][0.8\textwidth]
			- \big( \ddot z_0 \partial_\zeta + \ddot {\bar z}_0 \partial_{\bar\zeta} \big) \psi^{(n-1)}
			- \big( \dot z_0 \partial_\zeta + \dot {\bar z}_0 \partial_{\bar\zeta} \big) \tilde{\partial}_t \psi^{(n-1)} \\[0.2em]
			+ \big( \dot z_0 \partial_\zeta + \dot {\bar z}_0 \partial_{\bar\zeta}\big)^2 \psi^{(n-2)} 
			- r_q^{2\lambda} \exp(2\psi^{(0)}) \sum_{\substack{\pi \vdash n \\ \pi \neq [n]}} \prod_{i=1}^{|\pi|} \frac{1}{\nu_i!} \left(\psi^{(\pi_i)}\right)^{\nu_i}.
		\end{multlined}
\end{align}
Since $L_q$ is independent of $n$, the operators $\dKW_q^{(n)}$ only differ in the inhomogeneous terms $K_q^{(n)}$ determined by the lower order solutions $\psi^{(k)}$, $0\leq k\leq n-1$.
In particular, when $K_q^{(n)}$ is bounded each $\dKW_q^{(n)}$ is a Laplace-type operator and one can rely on the theory of elliptic operators.
More generally, since $\psi^{(0)}$ encodes Nahm pole boundary conditions and knot singularities, $\dKW_q^{(n)}$ is expected to be an iterated edge operator in the sense of \cite{Mazzeo1991, Mazzeo2013}.


\section{The Method of Continuity}
\label{sec:khovanov-method-of-continuity}

We propose that a continuity argument along the homotopy parameter $q\in[0,1]$ guarantees the existence of a solution of the decoupled Kapustin-Witten equations.
While a proof is currently out of reach, we sketch a strategy that offers potential avenues for further exploration.

Recall that our current goal is to determine a gauge transformation $g_\beta$ such that $g_\beta \cdot (A^\beta, \varphi^\beta, \phi_1^\beta)$ satisfies the decoupled Kapustin-Witten equations and exhibits Nahm pole boundary conditions as $y \to 0$ with knot singularities at $\beta_{q=1} = K$.
Using the homotopy expansion~\eqref{eq:khovanov-homotopy-expansion-of-dKW} of the decoupled Kapustin-Witten equations induced by the knot isotopy $\beta_\bullet$, it suffices to show that the set
\begin{align}
\mathcal{I} =\set{ q \in [0,1] \suchthat \exists\ \psi=\sum_{n=0}^\infty q^n \psi^{(n)}\ \text{ s.t. }\ \dKW_q^{(n)}(\psi^{(n)}) = 0,\ n \geq 0  } \subseteq [0,1].
\end{align}
is non-empty, open and closed.

In the following we lay out some initial considerations for each of these assertion.
Unfortunately, the proof strategy relies on analytic properties of the operators $\dKW_q^{(n)}$ that are currently not available and need a detailed investigation of their iterated edge structure.

\textbf{$\mathcal I$ is non-empty.}\quad
At $q=0$ the knot $\beta_0$ is the $S^1$-invariant single-stranded knot located at the origin of the complex plane.
In this case the equations reduce to $0=\dKW[q=0](\psi)=\EBE(\psi^{(0)})$, a solution of which is provided by the results of \cite{Mikhaylov2012}, so $0\in \mathcal{I}$.

\textbf{$\mathcal I$ is open.}\quad
Let $q_0\in \mathcal I$. 
We would like to show that there is an open neighbourhood $U_{q_0}$ of $q_0$ in $\mathcal{I}$.
For this, we first explain that in favourable circumstances there is such an open neighbourhood $U^{(n)}_{q_0}$ for each $\dKW_{q_0}^{(n)}$, individually.

Starting at $\mathcal{O}(q^0)$, let $\psi^{(0)}$ be a solution of $\EBE[q_0](\psi^{(0)}) = 0$.
This means that at each $t\in S^1_t$ it is given by~\eqref{eq:khovanov-knot-singularity-transformation} with $z$ replaced by $\zeta_{q_0}(t)$.
Explicitly, $\psi^{(0)} = \log \frac{\lambda+1}{R_{q_0}^{\lambda+1} s_{q_0} (s_{q_0})_\lambda}$, where $R_{q_0}$ and $s_{q_0}$ are spherical coordinates based at $(q_0 z_0(t),0) \in \mathbb{C}\times\mathbb{R}^+_y$.
If we replace $q_0$ in this expression by any other $q\in[0,1]$, the resulting function is again a solution of the extended Bogomolny equations, namely of $\EBE[q](\psi^{(0)}(\zeta_q(t),y)) = 0$.
It follows that $U^{(0)} = [0,1]$ is an open neighbourhood of $q_0$ for which there are solutions of $\EBE[q](\psi^{(0)}) = 0$, as requested.

Moving on to higher orders, let $n\geq 1$ and $(q_0, \psi^{(n)})$ a solution, i.e. $\dKW_{q_0}^{(n)}(\psi^{(n)}) = 0$.
Assume that the map $\dKW_{\bullet}^{(n)} : [0,1]\times \mathcal{X} \rightarrow \mathcal{Y}$ is Fr\'echet differentiable and that its linearization $\mathcal{L}_{(q_0,\psi^{(n)})}^{(n)}(0,-)$ at the point $(q_0, \psi_0^{(n)})$ is an isomorphism of Banach spaces (where $\mathcal{X},\mathcal{Y}$ are some appropriate function spaces that are attuned to Nahm pole boundary conditions with knot singularity).
The implicit function theorem then guarantees the existence of an open neighbourhood $U_{q_0}^{(n)}\subset[0,1]$ of $q_0$ and a function $G^{(n)}:U_{q_0}^{(n)}\rightarrow \mathcal{X}$, such that $\dKW_{q}^{(n)}(G^{(n)}(q)) = 0$ for all $q\in U_{q_0}^{(n)}$.

It then remains to show that the size of $U_{q_0}^{(n)}$ is bounded from below by some non-zero radius.
In that case the intersection $U_{q_0} := \bigcap_{n\in\mathbb{N}} U_{q_0}^{(n)}$ is open.
The size of the open neighbourhoods $U_{q_0}^{(n)}$ can be estimated from Lipschitz properties of the full Fr\'echet differential at $(q_0, \psi^{(0)})$.
Since the physical theory is topological, we can stretch the radius of $S^1_t$ and make $\abs{(d/dt)^n z_0}$ arbitrarily small.
One should expect that this can be leveraged to gain control over the Fr\'echet differential.
Once one has access to such bounds for each $\dKW_{q_0}^{(n)}$, one can conclude that for any point $q_0 \in \mathcal I$ the set $U_{q_0}$ is open.

By construction, for all $q\in U_{q_0}$ there is a sequence $\psi^{(n)}$ that satisfies $\dKW_q^{(n)}(\psi^{(n)})=0$.
In a last step, one needs to determine a function $\psi$ that has the formal power series $\psi = \sum q^n \psi^{(n)}$ near $q\in U$.
Recall that $\psi$ is assumed to be an adiabatic solution, in the sense that for small changes in time $t\to t+\epsilon$ the higher orders $\psi^{(n)}$ provide only miniscule adjustments that keep the configuration ``in equilibrium'', i.e. near a solution of the EBE.
This suggests that the corrections $\psi^{(n)}$ are small, and in particular small enough that the formal power series either converges or might be dealt with via Borel resummation for any $q$ that is close enough to $q_0$.

\textbf{$\mathcal I$ is closed.}\quad
Consider a sequence $\{q_k\} \subset \mathcal{I}$, together with a corresponding sequence $\{\psi_k\}$, such that $\dKW[q_k](\psi_k) = 0$.
$\{q_k\}$ converges in $[0,1]$ to some $q:=\lim_{k \to \infty} q_k$.
We need to show that a subsequence of $\psi_k$ converges to a corresponding $\psi$, such that $\dKW[q](\psi)$ vanishes to arbitrary order in $q$.

The desired statement would follow if we knew that the moduli space of solutions is compact.
Unfortunately, there is not yet a complete description of the moduli spaces of Kapustin-Witten solutions.
Although there has recently been progress for solutions on $X^3 \times \mathbb{R}^+$ when $X^3$ is compact \cite{Taubes2018}, the theory appears to involve a few subtleties that preclude a naive compactness result (also see related advances in \cite{Taubes2019, Taubes2021}).
Less is known about solutions of the decoupled Kapustin-Witten equations, but we expect that their Hermitian Yang-Mills structure provides some additional control.

In the situation relevant to us, it might be possible to directly construct a limiting solution, despite the lack of a general compactness theorem.
As before, this works almost trivially at order $q^0$.
To see this, consider for each $\psi_k$ the associated formal expansion $\psi_k = \sum q^n \psi_k^{(n)}$.
The terms at order $q^0$ define a sequence of EBE-solutions $\psi_k^{(0)} = \log \frac{\lambda+1}{R_{q_k}^{\lambda+1} s_{q_k} (s_{q_k})_\lambda}$.
This is continuous in the isotopy parameter, so in the limit $q_k \to q$ it approaches a limit $\psi^{(0)}$ where $q_k$ is replaced with $q$.
The limit then satisfies $\EBE[q](\psi^{(0)}) = 0$.

At order $q^1$ we are given the sequence of functions $\psi_k^{(1)}$ that each satisfy $\dKW_{q_k}^{(1)} \psi_k^{(1)} = 0$ with respect to their associated $q_k$.
Evaluating the action of $\dKW^{(1)}_q$ on each of the $\psi_k^{(1)}$ produces an error term
\begin{align}
	\dKW_q^{(1)}(\psi_k^{(1)}; \psi^{(0)}) = L_q \psi^{(1)} + K_q^{(1)}(\psi^{(0)}).
\end{align}
The analytic properties of the Laplace-type iterated edge operator $L_q$ are comparatively well-understood and it should be possible to utilize these to determine the size of the associated error term in relation to the distance $q-q_k$.
The formula for the inhomogeneities $K_q^{(1)}$ states that it is proportional to $\ddot z_0$.
It follows that $K_q^{(1)}$ is controlled by the magnitude of $\ddot z_0$, which can be made small by further stretching the knot in the direction of $t$.
Although we do not currently have a proof, we expect that one can construct an approximate solution $\psi^{(1)}$ from $\psi_k^{(1)}$ and improve it to arbitrary precision by taking $k\to \infty$.

Moving on from there, one can then proceed order by order in $q$ and in the end use a resummation argument to produce a solution $\psi = \sum q^n \psi^{(n)}$.


\section{Comoving Higgs Bundles}
\label{sec:khovanov-comoving-higgs-bundles}

In this section we move from a single-stranded knot in $S^1_t \times \mathbb{C}$ to the more general case of braids on $k$ strands in $S^1_t \times \Sigma$.
We find that the initial holomorphic data underlying the isotopy ansatz~\eqref{eq:khovanov-isotopy-ansatz} is captured by a one-parameter family of effective triples.
This is in analogy to the relation between the initial knot singularity ansatz~\eqref{eq:khovanov-knot-singularity-ansatz} and effective triples reviewed in \autoref{sec:khovanov-kobayashi-hitchin-correspondence}.

Let $K=\bigsqcup {}_{a=1}^k \{ (t, p_a(t), 0) \}$ be a braid in the boundary of $S^1_t \times \Sigma \times \mathbb{R}^+_y$.
The collection of trajectories $\{p_a(t)\}_{a=1,\ldots, k}$ can be viewed as the image of a map $\beta: S^1_t \to \Conf_k \Sigma$ in the configuration space of $k$ distinct, ordered points in $\Sigma$.
Equivalently, $\beta$ is an element of the loop space $\Omega \Conf_k \Sigma$.
The homotopy classes of loops form what is known as the pure braid group on $\Sigma$.

We are interested in the moduli space of solutions of the decoupled Kapustin-Witten equations.
\begin{definition}[Moduli Space of decoupled KW-solutions]
\label{def:khovanov-moduli-space-kapustin-witten}
\begin{align}
	\begin{multlined}[.8\linewidth]
	\widehat{\mathcal{M}}^{\mathrm{dKW}}_{K} := 
		\Big\{\
		\dKW(A,\phi) = 0 \mathrel{\Big|}
		(A, \phi) \text{ satisfies Nahm-pole boundary} \\
    \text{ conditions as } y\to 0
		\text{ with knot singularities at } K \\ 
		\text{ and converges to a flat } SL(n+1, \mathbb{C}) \text{ connection as } y \to \infty
		\Big\}.
		\end{multlined}
\end{align}
\end{definition}
As before, we denote by $\mathcal{M}^{\mathrm{dKW}}_K$ the quotient by real gauge transformations $\mathcal{G}_0(S^1_t\times \Sigma \times \mathbb{R}^+_y)$ that vanish at the boundary.
However, most of the upcoming discussions will focus on the infinite dimensional moduli spaces before modding out gauge transformations.

Let us slightly change perspective and view the knot singularity data as part of the moduli of the problem.
Assume, for the sake of simplicity, that all strands of $K$ are labeled by the same weight $\lambda \in \Gamma_{\mathrm{char}}^\vee$.
Specifically, we will from now on view the moduli spaces $\widehat{\mathcal{M}}^{\mathrm{dKW}}_K$ as fibers of the following bundle:
\begin{align}
\begin{tikzcd}
	\widehat{\mathcal{M}}^{\mathrm{dKW}} \arrow{d} \\ 
	\Omega \Conf_k \Sigma
\end{tikzcd}
\label{eq:khovanov-dKW-fibration}
\end{align}
The fiber map sends each solution to the loop $\beta\in\Omega\Conf_k\Sigma$ along which the solution exhibits a knot singularity.

There is an analogous fiber bundle $\widehat{\mathcal{M}}^{\mathrm{EBE}} \to \Conf_k \Sigma$ for solutions of the extended Bogomolny equations.
In this case the fibers are given by $\widehat{\mathcal{M}}^{\mathrm{EBE}}_D$ with fixed knot singularity data $D = \{ (p_a, \lambda_a) \}_{a=1,\ldots, k}$ and the fiber map sends a given solution to $\{p_a\}_{a=1,\ldots,k} \subset \Sigma$.

Recall from \autoref{sec:khovanov-kobayashi-hitchin-correspondence} that the moduli spaces $\mathcal{M}^{\mathrm{EBE}}_D$ are in bijection with the moduli space of effective triples $\mathcal{M}_{(\mathcal{E},\varphi,L)}$, where the associated divisor $\mathfrak{d}(\varphi,L)$ was required to coincide with the knot singularity data $D$.
Reinterpreting the data of the divisor as part of the moduli of an effective triple, we obtain a corresponding fiber bundle:
\begin{align}
\begin{tikzcd}
	\widehat{\mathcal{M}}_{(\mathcal{E},\varphi,L)} \arrow{d} \\ 
	\Conf_k \Sigma
\end{tikzcd}
\label{eq:khovanov-EBE-fibration}
\end{align}
The fiber map sends each effective triple to the points of its divisor $\mathfrak{d}(L,\varphi) = \{ (p_a, \lambda_a) \}_{a=1,\ldots k}$.

\begin{remark}
More generally, if one includes the information of weights $\lambda_a$ at each point, the base of these fibrations is the configuration space of labeled points $\Conf_{k}( \Sigma ;\lambda_1, \ldots, \lambda_k )$ and its corresponding loop space, respectively.
\end{remark}

The same arguments as in \autoref{sec:khovanov-kobayashi-hitchin-correspondence} can be used to show that any solution $(A,\phi)$ of the decoupled Kapustin-Witten equations gives rise to a one-parameter family of ``comoving'' effective triples $\{ (\mathcal{E}_t, \varphi_t, L_t) \}_{t\in S^1}$.
To see this, assume $(A,\phi)$ is a solution of the decoupled Kapustin-Witten equations with knot singularity along a braid $\beta = \bigsqcup \{p_a(t)\}$.
Let $\mathcal{D}_\mu$ denote the operators associated to $(A,B)$ and work in temporal gauge $A_s=0$, such that $\mathcal{D}_0 = i D_t$.

The decoupled Kapustin-Witten equations contain the equations $[\mathcal{D}_i, \mathcal{D}_j]= 0$, $i,j=1,2,3$.
Denote by $V$ the vector bundle associated to the fundamental representation of $SL(N,\mathbb{C})$ and let $V_{(t,y)}$ be its restriction to $\{t\} \times \Sigma \times \{y\}$.
As before, $\mathcal{D}_1$ provides a $\bar\partial$ operator, making $V_{(t,y)}$ into a holomorphic vector bundle $\mathcal{E}_{(t,y)}$, while $\mathcal{D}_3 = \ad_\varphi$ is a holomorphic $K_\Sigma$-valued endomorphism of $\mathcal{E}_{(t,y)}$.
The associated family of Higgs bundles $(\mathcal{E}_{(t,y)}, \varphi_{(t,y)})$ over $\Sigma$ is parallel with respect to $\mathcal{D}_2$, such that for each $t\in S^1$ the Higgs bundle at $y$ is determined via parallel transport of a stable reference Higgs bundle $(\mathcal{E}_t,\varphi_t)$ that sits, for example, at $y=\infty$.
Finally, for each $t\in S^1$, the boundary condition at $y\to 0$ determines a vanishing line bundle $L_t \subset \mathcal{E}_t$ whose divisor $\mathfrak{d}(L,\varphi) = \{p_a(t), \lambda\}$.

The decoupled Kapustin-Witten equations additionally include the operator $\mathcal{D}_0$, which yields a notion of parallel transport along $S^1_t$.
The equations $[\mathcal{D}_0,\mathcal{D}_i] = 0$ state that $(\mathcal{E}_t, \varphi_t, L_t)$ is parallel.
Equivalently, parallel transport via $\mathcal{D}_0$ defines an Ehresmann connection on $\widehat{\mathcal{M}}_{(\mathcal{E},\varphi,L)}$ with respect to which $(\mathcal{E}_t, \varphi_t, L_t)$ is a horizontal lift of the braid $\beta$.

Denote by $\Omega_h \widehat{\mathcal{M}}_{(\mathcal{E},\varphi,L)}$ the space of non-vertical loops, i.e. those loops that are horizontal with respect to \emph{some} Ehresmann connection.
The discussion above implies that there is the following Kobayashi-Hitchin-like bundle map:
\begin{equation}
	\label{eq:khovanov-KH-fibration}
	\begin{tikzcd}
		\widehat{\mathcal{M}}^\mathrm{dKW}  \arrow[rr, "\hat{I}_{\mathrm{KH}}"] \arrow[rd] &[-30pt] 
		&[-30pt] 
		\Omega_h \widehat{\mathcal{M}}_{(\mathcal{E},\varphi,L)} \arrow[dl] 
		\\
		&  
		\Omega \Conf_k \Sigma &
		&
	\end{tikzcd}
\end{equation}
We expect that this descends to a corresponding map on quotient spaces after modding out gauge transformations:
\begin{align}
	I_{\mathrm{KH}}: \mathcal{M}^{\mathrm{dKW}} \to \Omega_h \widehat{\mathcal{M}}_{(\mathcal{E},\varphi,L)} / \mathcal{G}_{\mathbb{C}}(S^1_t \times \Sigma).
\end{align}
Note that this Kobayashi-Hitchin bundle map contains He and Mazzeo's classification of $S^1$-invariant Ka\-pu\-stin-\-Wit\-ten solutions in terms of EBE-solutions (\autoref{thm:khovanov-He-Mazzeo-S1-invariant-classification}).
Keeping in mind that according to \autoref{thm:khovanov-Gaiotto-Witten-bijections} there is a bijection $I_{NPK}: \mathcal{M}^{\mathrm{EBE}} \to \mathcal{M}_{(\mathcal{E},\varphi, L)}$, their result states that there is a bijection of fibers over the constant loops in $\Omega \Conf_k \Sigma$:
\begin{center}
\begin{tikzcd}
	\mathcal{M}^\mathrm{dKW}  \arrow[r, "I_{\mathrm{KH}}"] \arrow[rd] \arrow[rr, bend left, "I_{NPK}"] &[1em]
	\Omega_h \mathcal{M}_{(\mathcal{E},\varphi,L)} \arrow[d] \arrow[r, hookleftarrow, "\mathrm{const.}"'] &[1em]
	\mathcal{M}_{(\mathcal{E},\varphi,L)} \arrow[d] \\
	&[1em]
	\Omega \Conf_k \Sigma \arrow[r, hookleftarrow, "\mathrm{const.}"] &[1em]
	\Conf_k \Sigma
\end{tikzcd}
\end{center}
Put differently, the map $I_{\mathrm{KH}}$ is expected to be a generalization of the map $I_{NPK}$ to the case of non-vertical loops in the moduli space of decoupled Kapustin-Witten solutions.

As a next step, we describe how to obtain a solution of the decoupled Kapustin-Witten equations from a family of effective triples.
Hence, suppose we are given a non-vertical loop of effective triples $\gamma(t) = (\mathcal{E}(t), \varphi(t), L(t))$.
The knot singularity data associated to $\gamma(t)$ under the fiber map is a braid $\beta(t) = \{ p_a(t) \}_{a=1,\ldots, k}$.
We wish to construct from $\gamma$ a field configuration $(A,B)$ that exhibits a Nahm pole with knot singularities along $\beta$ and satisfies the decoupled Kapustin-Witten equations.

\begin{figure}
\centering
\includegraphics[width=\textwidth, height=0.35\textheight, keepaspectratio]{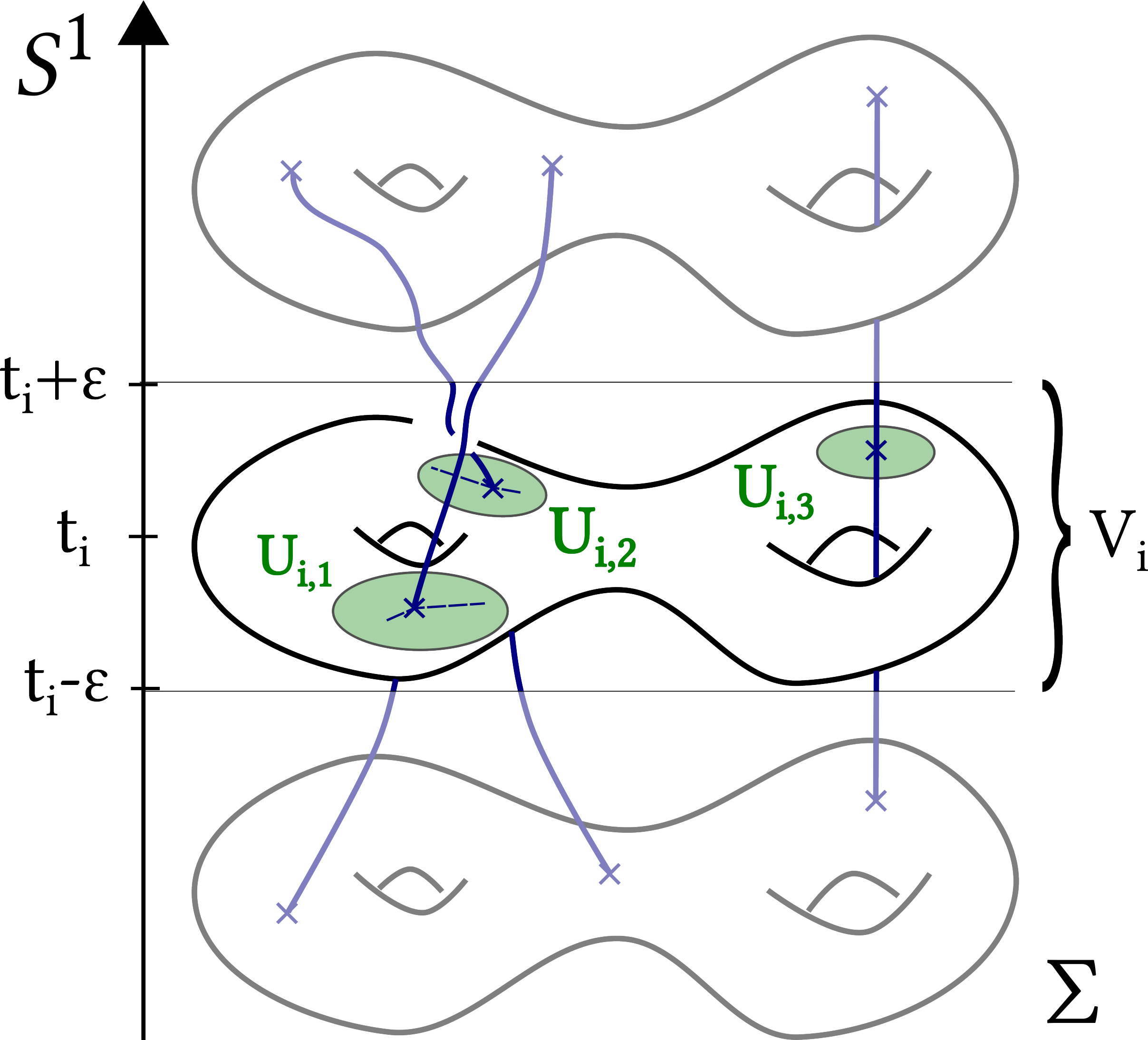}
\caption{Illustration of the covering of $S^1_t \times \Sigma \times \mathbb{R}^+_y$ that is used in the iterative construction of approximate solutions of the decoupled Kapustin-Witten equations for a multi-stranded and time-dependent knot.}
\label{fig:khovanov-cover-braid}
\end{figure}

Under the assumption that the continuity method of \autoref{sec:khovanov-method-of-continuity} provides a solution for any single-stranded knot in $\mathbb{C}$, this can be achieved by essentially the same arguments as in the context of the Kobayashi-Hitchin correspondence for EBE-solutions described in \autoref{sec:khovanov-kobayashi-hitchin-correspondence}.
To that end, cover the manifold by open slices $V_i = (t_i-\epsilon,t_i+\epsilon) \times \Sigma \times \mathbb{R}_+$ and write $\beta_i = \{ p_{i,a}(t)\}_{a=1,\ldots k}$ for the part of the braid that lies in $V_i$, see \autoref{fig:khovanov-cover-braid}.
Choose the slices $V_i$ thin enough that there exists an isotopy that interpolates between $\beta_i$ and some $S^1$-invariant braid $\{P_{i,a} = \text{const.} \}_{a=1,\ldots,k} \in \Conf_k \Sigma$.
Moreover, we demand that this isotopy is of a particularly mild form, namely such that there exist non-intersecting discs $U_{i,a}$, centered at $P_{i,a}$, which each contain the corresponding trajectory $p_{i,a}(t)$ for all $t \in (t_i - \epsilon, t_i + \epsilon)$ and the isotopy that connects $p_{i,a}(t)$ to $P_{i,a}$.

We construct an approximate solution of the decoupled Kapustin-Witten equations according to the following outline.
Starting with a single slice $V_i$, choose an open cover of $\Sigma$ consisting of an open set $U_{i,0}$ that does not contain any of the trajectories $p_{i,a}(t)$, together with the open discs $U_{i,a}$ described above.
Using the method of continuity of \autoref{sec:khovanov-method-of-continuity}, we know that there is a field configuration that satisfies Nahm pole boundary conditions with knot singularity on each $(t_i-\epsilon, t_{i}+\epsilon) \times U_{i,a} \times \mathbb{R}^+_y$ independently.
These configurations are then glued ``horizontally'' over $U_{i,0}$ to an approximate solution on $V_i$, which subsequently can be glued ``vertically'' over the intersections of $V_i \cap V_{i+1}$ to produce an approximate solution on $S^1 \times \Sigma \times \mathbb{R}^+_y$.
The construction proceeds in five steps.

First, dropping the indices $i$ and $a$ for the moment, consider a small disk $U$ centered at $P$ and containing the trajectory $p(t)$ of a single strand.
Let $z$ be a holomorphic coordinate on $U$, assume $P$ corresponds to $z=0$, and denote the coordinates of $p(t)$ by $z_0(t)$.
We also use coordinates $(t,R,\psi,\vartheta)$, with spherical coordinates on the second and third factor of $S^1_t \times U \times \mathbb{R}^+_y$.
We may then extract from $\gamma(t)=(\mathcal{E}(t),\varphi(t),L(t))$ a field configuration on $[t_i-\epsilon, t_{i}+\epsilon] \times U \times \mathbb{R}^+_y$ that looks like the isotopy ansatz $(A^\beta,\varphi^\beta,\phi_1^\beta)$ as given in~\eqref{eq:khovanov-isotopy-ansatz}:
Recall that for each $t\in (t_i-\epsilon, t_i + \epsilon)$, the effective triple defines a frame $\{ e_1, \ldots e_{n+1}\}$ of $\mathcal{E}_t$ with respect to which the Higgs field is of the form $\varphi(t) = \sum (z-z_a(t))^{\lambda_i} E_i^+\ dz$.
The isotopy that connects $P$ to $p(t)$, and which is contained in $U$ by assumption, is homotopy equivalent to the isotopy given by $\zeta_q(t) = z - qz_a(t)$.
Replacing $z-z_a(t)$ by $\zeta_q(t)$ in the expression for $\varphi(t)$ then provides the $q^0$ part of the isotopy ansatz.
To get the part proportional to $q^1$, we identify the gauge field component $A^\beta_t$ with the connection form of some Ehresmann connection with respect to which $\gamma$ is horizontal.
Since $D_t \varphi(t) = 0$, the connection form satisfies $[A_t, \varphi(t)] = -i \partial_t \varphi(t)$.
This is solved by setting $A^\beta_t = \frac{q \dot z_a}{\zeta} \sum \lambda_i A_{ij}^{-1} H_j$.

In the second step, we rely on the method of continuity to invoke the existence of a complex gauge transformation $g_U$ that maps the field configuration $(A^\beta, \varphi^\beta, \phi_1^\beta)$ to an approximate solution of the decoupled Kapustin-Witten equations on $(t_i-\epsilon, t_i + \epsilon) \times U \times \mathbb{R}^+_y$.
By construction, this satisfies the Nahm pole boundary conditions with knot singularity at $p(t)$.

The third step constructs a gauge transformation $g_{V_i}$ on $V_i = (t_i-\epsilon, t_i + \epsilon) \times \Sigma \times \mathbb{R}^+_y$ from the collection of local frames and gauge transformations $\{ g_{\lambda_a} \}$ associated to $U_a$.
To do so, we additionally choose a holomorphic frame on $(t_i-\epsilon, t_i + \epsilon) \times U_0 \times \mathbb{R}^+_y$ and let $g_0$ be the gauge transformation defined in~\eqref{eq:khovanov-nahm-pole-singularity-transformation}.
Just as before, on the overlaps $U_0 \cap U_a$, the holomorphic frames are related by the transition functions
\begin{align}
	g_{0a} = \exp\big( - \log r \sum_{i=1}^n  \lambda_{a,j} A_{ij}^{-1}  H_i \big).
\end{align}
Gluing the gauge transformations $g_0$ and $g_{U_a}$ via a partition of unity produces the desired $g_{V_i}$.

Fourth, the holomorphic frames on $U_{i,0}$ and $U_{i+1,0}$ are equivalent up to gauge transformations, so they can in turn can be glued with $g_{V_{i+1}}$ on the intersection $V_i \cap V_{i+1}$.

The resulting field configuration is a first approximation to a solution of the decoupled Ka\-pu\-stin-\-Wit\-ten equations.
It is of order $\mathcal{O}(y^{-1})$ away from the braid $\beta$ and of order $\mathcal{O}(R^{-1}s^{-1})$ near any of its strands, which means that it satisfies Nahm pole boundary conditions with knot singularities.
The approximation can be improved order by order, first to desired precision in $R$ near any point $p\in K$ and afterwards in~$y$.

\section{Effective Triples, Monopoles, and the Grothendieck-Springer Fibration}
\label{sec:khovanov-effective-triples-and-monopoles}

Unfortunately, the analytic properties of the Haydys-Witten and Kapustin-Witten equations and their moduli spaces are not particularly well understood.
We propose that when $\Sigma = \mathbb{C}$, there is a finite-dimensional fiber bundle that replaces the unknown fiber bundle in~\eqref{eq:khovanov-KH-fibration} and encodes enough information to provide existence results for solutions of the decoupled Kapustin-Witten equations.
The fiber bundle in question is a variant of the Grothendieck-Springer resolution of $\mathfrak{sl}(kN,\mathbb{C})$, which also appears in the definition of symplectic Khovanov homology \cite{Seidel2004, Manolescu2007}.
The motivation to consider this space as a finite-dimensional model of the problem is ultimately based in physical properties of the system and motivated by observations of Gaiotto and Witten \cite{Gaiotto2012a}.
In some sense this is comparable to an ADHM-like approach, where the construction of Yang-Mills instantons is reduced to a finite dimensional problem in linear algebra \cite{Atiyah1978}.

Consider the image of a braid $\beta$ on $k$ strands in $S^1_t \times \mathbb{C}$ and denote the trajectories of the strands in $\mathbb{C}$ by $z_a(t)$.
Our interest is in the time evolution of $k$-monopole solutions of the extended Bogomolny equations along $\beta$.
For this, we view the monopole insertions at $(t,z_a(t),0)$ in $S^1_t \times \mathbb{C} \times \mathbb{R}^+_y$ as ``heavy particles'' with magnetic charge $\lambda_a$, that trace out fixed, non-colliding trajectories in $\mathbb{C}$.
Since the underlying physical theory is topological, these particles don't interact and the multi-particle system is equivalent to the union of $k$ single particles.
Correspondingly, the configuration of each individual monopole is fully determined in a small neighbourhood.
The assumption that the particles are ``heavy'' means that the evolution of each monopole along $S^1_t$ is viewed as an externally determined background in which the quantum theory lives.
More specifically, the quantum system starts at an initial time in some ground state and then evolves adiabatically along $S^1_t$, meaning that at any given time $t$ the system remains in a ground state of the corresponding, fixed background configuration at that time.
This picture suggests that, as a first step, we need to determine the moduli of a collection of $k$ individual monopoles.

According to the Kobayashi-Hitchin correspondence discussed in \autoref{sec:khovanov-kobayashi-hitchin-correspondence}, a $k$-monopole solution of the extended Bogomolny equations is determined by an effective triple.
Hence, assume we are given $(\mathcal{E}, \varphi, L)$ with associated divisor $\mathfrak{d}(L,\varphi) = \{(z_a, \lambda_a)\}_{a=1,\ldots, k}$.
Due to the physical argument above, we now restrict our attention to the moduli of the Higgs field $\varphi$ near the points of $\mathfrak{d}(L,\varphi)$.
We have seen in \autoref{sec:khovanov-kobayashi-hitchin-correspondence} and \autoref{sec:khovanov-comoving-higgs-bundles} that on a small enough disc $U_a \subset \mathbb{C}$ centered at $z_a$ the pair $(L,\varphi)$ is equivalent to the data contained in the ansatz $\varphi^\lambda\restr_{U_a} = \sum_i z^{\lambda_{a,i}} E_i^+$.
The latter, in turn, is determined by the choice of
\begin{enumerate}
\item a nilpotent element $E_a \in \mathcal{O}_{\pi_a}$, such that $\varphi^\lambda{}\restr_{z=0} = E_a$, and
\item a nilpotent element $K_a \in \ker{F_a}$, such that $\varphi^\lambda{}\restr_{z=1} = E_a + K_a \in \mathcal{O}_{\mathrm{reg}}$.
\end{enumerate}
Recall from \autoref{sec:khovanov-Nahm-pole-boundary-condition} that the partition $\pi_a = [\pi_{a,1} \ldots \pi_{a,s}]$ is determined by the monopole charge $\lambda_a$ and given by counting the numbers of consecutive Dynkin labels in $\lambda_a = (\lambda_{a,1}, \ldots, \lambda_{a,n})$ that vanish.
In conclusion, individual monopoles are determined by a choice of element in the intersection of the Slodowy slice $\mathcal{S}_{E_a}$ and the regular nilpotent orbit $\mathcal{O}_{\mathrm{reg}}$ in $\mathfrak{sl}(N,\mathbb{C})$.

A naive approach to keep book of all $k$ monopoles simultaneously is to combine their moduli into a coproduct.
In finite dimensions this amounts to putting everything together into a single matrix $Y$ of size $kN$:
\begin{align}
	Y
	:= \left(
		\begin{array}{c|c|c}
			E_{1} + K_1 & 0 & 0 \\ \hline
			0 & \ddots & 0 	 	 \\ \hline
			0 & 0 & E_{k} + K_k \\
		\end{array}
		\right).
\end{align}
Put differently, $Y$ is an element of
\begin{align}
		(\mathcal{S}_{E_1} \cap \mathcal{O}_{\mathrm{reg}}) \oplus \ldots \oplus (\mathcal{S}_{E_k} \cap \mathcal{O}_{\mathrm{reg}})
		\subset \mathfrak{sl}(kN,\mathbb{C}).
\end{align}
Remember that $\mathcal{O}_{\mathrm{reg}} = \mathcal{O}_{[N]}$ and let $\rho := [N^k]$ be the partition of $kN$ that consists of $k$ copies of $N$.
Similarly, denote the partition of $kN$ that is obtained from concatenating all $\pi_a$ by $\pi = [\pi_1 \ldots \pi_k]$.
There exists a $g\in SL(kN,\mathbb{C})$ that maps $E_1 \oplus \ldots \oplus E_k$ to its Jordan normal form $E_\pi$.
It follows that the $k$-monopole configuration $Y$ is always conjugate to an element of $\mathcal{S}_{\pi} \cap \overline{\mathcal{O}_\rho}$ (recall $\mathcal{S}_\pi := \mathcal{S}_{E_\pi}$).
The upcoming constructions only depend on the topology and geometry of Slodowy slices, which are independent of the base point.
Accordingly, we from now on consider a $k$-monopole configuration to be determined by a choice of
\begin{align}
		Y \in \mathcal{S}_\pi \cap \overline{\mathcal{O}_{\rho}} \subset \mathfrak{sl}(kN,\mathbb{C}).
	\label{eq:khovanov-naive-model-space}
\end{align}

Unfortunately, there are fundamental problems in determining Kapustin-Witten solutions from nilpotent Higgs bundles directly \cite{Gaiotto2012a, Dimakis2022a, Sun2023}.
Inspired by the ``complex symmetry breaking'' suggested by Gaiotto and Witten to circumvent these problems, and also because we wish to encode the positions $z_a$ of the monopoles in a fiber bundle structure, we modify the naive approach.
Namely, we additionally encode the position $z_a$ of each strand by a deformation of the nilpotent orbit $\overline{\mathcal{O}_\rho}$ in~\eqref{eq:khovanov-naive-model-space}.

Define the traceless, diagonal $N\times N$-matrix
\begin{align}
	D(z_a) = \diag\big( z_a,\ \ldots,\ z_a,\ -(N-1) z_a \big) \in \mathfrak{sl}(N,\mathbb{C}).
\end{align}
We call $z_a$ ``thick'' eigenvalue of $D(z_a)$ as it appears with multiplicity $N-1$, while we refer to the remaining eigenvalue $-(N-1)z_a$ as ``thin''.

For fixed $t\in S^1_t$, denote the position of the strands by $D = \{ z_a(t) \}_{a=1,\ldots, k} \in \Conf_k \mathbb{C}$.
Write $\mathcal{O}_{\rho, D}$ for the adjoint orbit of $D(z_1) \oplus \ldots \oplus D(z_k)$ in $\mathfrak{sl}(kN,\mathbb{C})$ and consider the map
\begin{align}
	Y &\mapsto Y + \left( D(z_1) \oplus \ldots \oplus D(z_k) \right).
\end{align}
The eigenvalues of this new matrix are given by the positions $z_a$, each with (generalized) eigenspace of dimension $(N-1)$, and the values $-(N-1)z_a$ with eigenspace of dimension $1$.
This defines a smooth deformation $\overline{\mathcal{O}_{\rho}} \leadsto \mathcal{O}_{\rho, D}$.

We conclude that the moduli space of $k$ monopoles of charge $\lambda$ and located at $D=\{ z_a\}$ is given by
\begin{align}
	\mathcal{Y}_{\pi,\rho,D} := \mathcal{S}_{\pi} \cap \mathcal{O}_{\rho, D} \subset \mathfrak{sl}(kN, \mathbb{C}).
\end{align}
As an intersection of a Slodowy slice with an adjoint orbit, $\mathcal{Y}_{\pi,\rho,D}$ is a finite dimensional K\"ahler manifold.

\begin{remark}
Going from $\widehat{\mathcal{M}}_{(\mathcal{E},\varphi,L)} \to \mathcal{Y}_{\pi, \rho, D}$ can be viewed as modding out a class of ``large'' gauge transformations that are non-zero at the boundary.
A priori, we are not allowed to mod out such gauge transformations, because the associated boundary conditions are physically distinguishable.
However, there is some evidence that for the class of large gauge transformations in question there is always a Haydys-Witten instanton that interpolates between the associated two (a priori inequivalent) solutions of the decoupled Haydys-Witten equations.
Such a result would provide a rigorous a posteriori justification for replacing the infinite dimensional moduli space of effective triples by $\mathcal{Y}_{\pi,\rho,D}$.
We further comment on this in \autoref{sec:khovanov-homology}.
\end{remark}

A variant of the space $\mathcal{Y}_{\pi,\rho,D}$ plays a major role in the definition of symplectic Khovanov homology and symplectic $\mathfrak{sl}(N,\mathbb{C})$-Khovanov-Rozansky homology \cite{Seidel2004, Manolescu2007}.
Importantly, $\mathcal{Y}_{\pi,\rho,D}$ can be viewed as a fiber of what Manolescu calls a \emph{restricted partial simultaneous Grothendieck resolution of the adjoint quotient map}.
For our purposes it will suffice to define the relevant version of the adjoint quotient map ``by hand'' and we refer to \cite{Manolescu2007} for a more satisfactory exposition of the general construction.

Remember that the adjoint quotient map $\chi : \mathfrak{sl}(kN, \mathbb{C}) \rightarrow \mathfrak{h}/\mathcal{W}$ sends a matrix to its generalized eigenvalues.
Denote by $\mathfrak{sl}(kN, \mathbb{C})^{[(N-1)^k 1^k]}$ the space of traceless $kN \times kN$ matrices that have $k$ pairs of eigenvalues that cancel each other in the trace, where each pair has one ``thick'' eigenvalue of algebraic multiplicity $N-1$ and one ``thin'' eigenvalue of multiplicity $1$, respectively.
A prototypical element of this space is $D(z_1)\oplus\ldots\oplus D(z_k)$, but we explicitly allow non-diagonalizable elements with Jordan blocks of size greater than 1.
Define $\tilde\chi$ to be the map that sends an element of $\mathfrak{sl}(kN, \mathbb{C})^{[(N-1)^k 1^k]}$ to its thick eigenvalues $(z_1, \ldots, z_k) \in \Conf_k \mathbb{C}$:
\begin{align}
	\tilde\chi : \mathfrak{sl}(kN,\mathbb{C})^{[(N-1)^k\; 1^k]} \to \Conf_k \mathbb{C}.
\end{align}
We will simply refer to this as Grothendieck-Springer fibration.
With that understood $\mathcal{Y}_{\pi,\rho,D}$ is equivalent to the following fiber of the restriction of $\tilde \chi$ to the Slodowy slice $S_{\pi}$:
\begin{align}
	\mathcal{Y}_{\pi,\rho,D} = \tilde\chi \restr_{S_{\pi}}^{-1}\big( D(z_1) \oplus \ldots \oplus D(z_k) \big).
\end{align}

Our discussion so far shows that every effective triple $(\mathcal{E},\varphi,L)$ with divisor $\mathfrak{d}(L,\varphi) = D$ determines an element $Y \in \mathcal{Y}_{\pi,\rho,D}$.
This induces the following bundle map:
\begin{align}
\begin{tikzcd}[ampersand replacement=\&]
	\widehat{\mathcal{M}}_{(\mathcal{E},\varphi,L)}  \arrow[rr, "\varUpsilon"] \arrow[rd]  \&
	\&
	 S_\pi \cap \mathfrak{sl}(kN,\mathbb{C})^{[(N-1)^k\; 1^k]}\arrow[dl, "\tilde\chi\restr_{S_\pi}"] 
	\\
	\& 
	\Conf_k \mathbb{C} \&
\end{tikzcd}
\label{eq:khovanov-effective-triple-to-monopole-moduli}
\end{align}
$\varUpsilon$ is clearly not injective, since it maps effective triples that differ in regions of $\Sigma$ sufficiently far from any monopole insertions to the same element in $\mathcal{Y}_{\pi,\rho,D}$.
However, we expect that $\varUpsilon$ is surjective.
If this holds, we are guaranteed that for every non-vertical path in $\mathfrak{sl}(kN,\mathbb{C})^{[(N-1)^k\; 1^k]}$, there exists at least one non-vertical path in $\widehat{\mathcal{M}}_{(\mathcal{E},\varphi,L)}$, which in turn determines a solution of the decoupled Kapustin-Witten equations by the arguments presented in \autoref{sec:khovanov-comoving-higgs-bundles}.

\begin{figure}
	\centering
	\includegraphics[width=\textwidth, height=0.3\textheight, keepaspectratio]{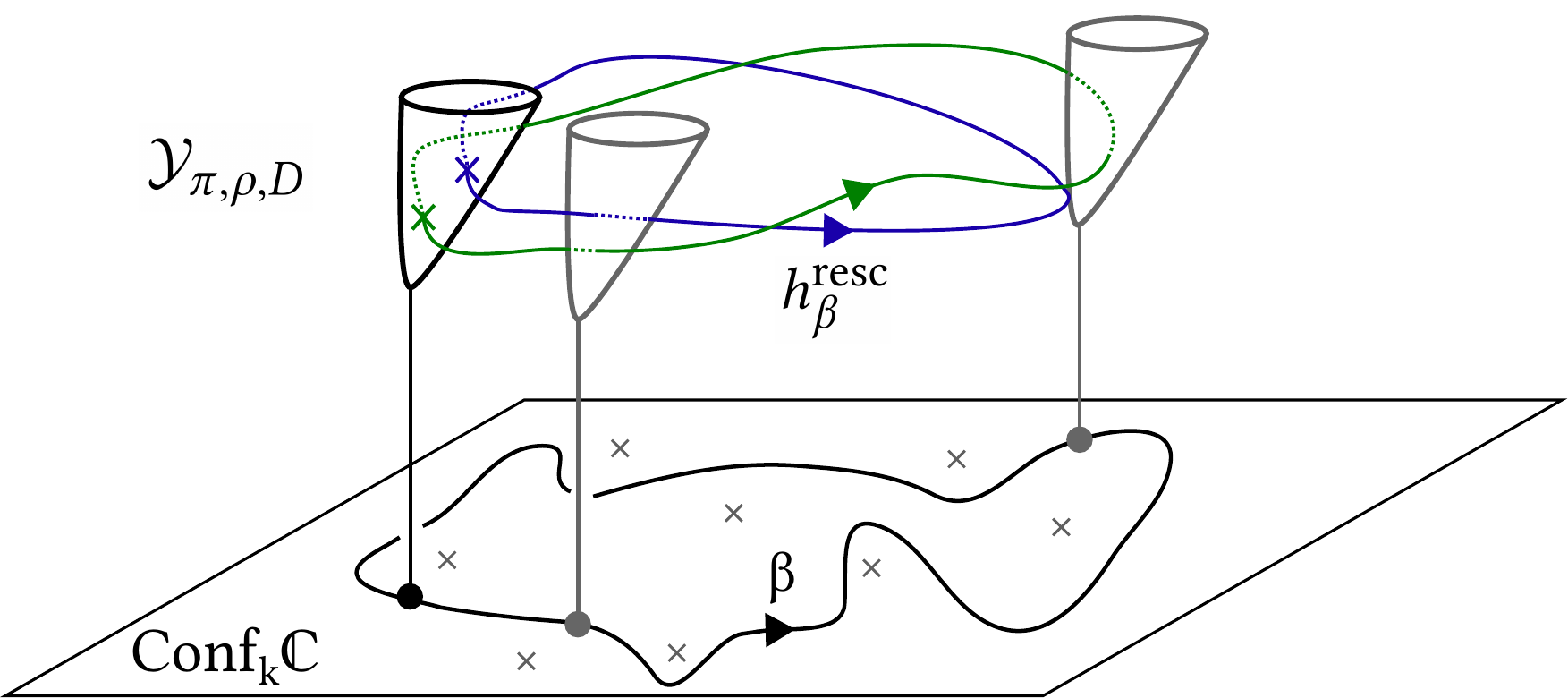}
	\caption{The fixed points of rescaled parallel transport $h_\beta^{\mathrm{resc}}: \mathcal{Y}_{\pi,\rho,D} \rightarrow \mathcal{Y}_{\pi,\rho,D}$ along a pure braid $\beta$ determine inequivalent horizontal lifts of~$\beta$. Crosses in the }
	\label{fig:khovanov-parallel-transport-in-fiber}
\end{figure}
	
As discussed in \cite[Sec. 4A]{Seidel2004} and \cite[Sec. 4.1]{Manolescu2007}, since $\mathfrak{sl}(kN,\mathbb{C})^{[(N-1)^k\; 1^k]}$ carries a K\"ahler metric, there exists a suitable notion of (rescaled) parallel transport along paths $\beta:[0,1]\rightarrow \Conf_k \mathbb{C}$ from $D=\beta(0)$ to $D^\prime=\beta(1)$.
\begin{align}
	h_\beta^{\mathrm{resc}}: \mathcal{Y}_{\pi,\rho,D} \rightarrow \mathcal{Y}_{\pi,\rho,D^\prime}.
\end{align}
Some care is needed in defining $h_\beta^{\mathrm{resc}}$, since $\mathfrak{sl}(kN,\mathbb{C})^{[(N-1)^k\; 1^k]}$ is not compact and $\tilde\chi$ may fail to be a proper map.
In that situation the integral lines of the vector field $H_\beta = \dot\beta\ \nabla \tilde\chi / \norm{\nabla \tilde\chi}^2 $, which is parallel to $\beta$ and orthogonal to the fibers, may not exist at all times.
To circumvent this problem, one rescales $H_\beta$ in such a way that its integral lines stay in some controlled compact subset of $\mathfrak{sl}(kN,\mathbb{C})^{[(N-1)^k\; 1^k]}$.
Strictly speaking, $h_\beta^{\mathrm{resc}}$ is only well-defined on (arbitrarily large) compact subsets.

We can apply parallel transport along a pure braid $\beta: S^1 \to \Conf_k \mathbb{C}$, in which case $\mathcal{Y}_{\pi,\rho,D} = \mathcal{Y}_{\pi,\rho,D^\prime}$.
Horizontal lifts of $\beta$ that start and end at the same point in $\mathcal{Y}_{\pi,\rho,D}$ are then in one-to-one correspondence with fixed points of $h_\beta^{\mathrm{resc}}$, see \autoref{fig:khovanov-parallel-transport-in-fiber}.
With this we arrive at one of the main claims of this article:
\begin{conjecture}
\label{conj:lower-bound-for-KW-solns}
The number of solutions to the decoupled Kapustin-Witten equations on $S^1_t \times \mathbb{C} \times \mathbb{R}^+_y$ with knot singularities along $\beta$ of weight $\lambda$ is bounded from below by the number of fixed points of $h_\beta^{\mathrm{resc}}$.
\end{conjecture}
In the upcoming sections the ideas of this section are applied to a slightly modified setting, where $S^1_t$ is decompactified to $\mathbb{R}_t$ and the braid is replaced by a compact knot $K$.
This will lead us to a considerably stronger version of \autoref{conj:lower-bound-for-KW-solns}, providing a version of Witten's conjecture that relates Haydys-Witten instanton homology to symplectic Khovanov homology.


\section{From Braids to Knots}
\label{sec:khovanov-braids-to-knots}

We now apply the ideas of the previous section to the decoupled Kapustin-Witten equations with knot singularities along a compact knot $K$ in the boundary of $W^4 = \mathbb{R}_t \times \mathbb{C} \times \mathbb{R}^+_y$.
Since the factor of $S^1_t$ is decompactified, we can no longer identify $K$ with an element of $\Omega \Conf_k \mathbb{C}$.
However, by Alexander's theorem any knot $K$ can be represented as the closure of a braid $\beta$ on $k$ strands.
Equivalently, $K$ can be identified with a particular loop in the closure of the configuration space $\overline{\Conf_k \mathbb{C}}$ that starts and ends at singular strata as explained below.

\begin{figure}
\centering
\begin{minipage}{5cm}
\includegraphics[width=5cm, height=5cm, keepaspectratio]{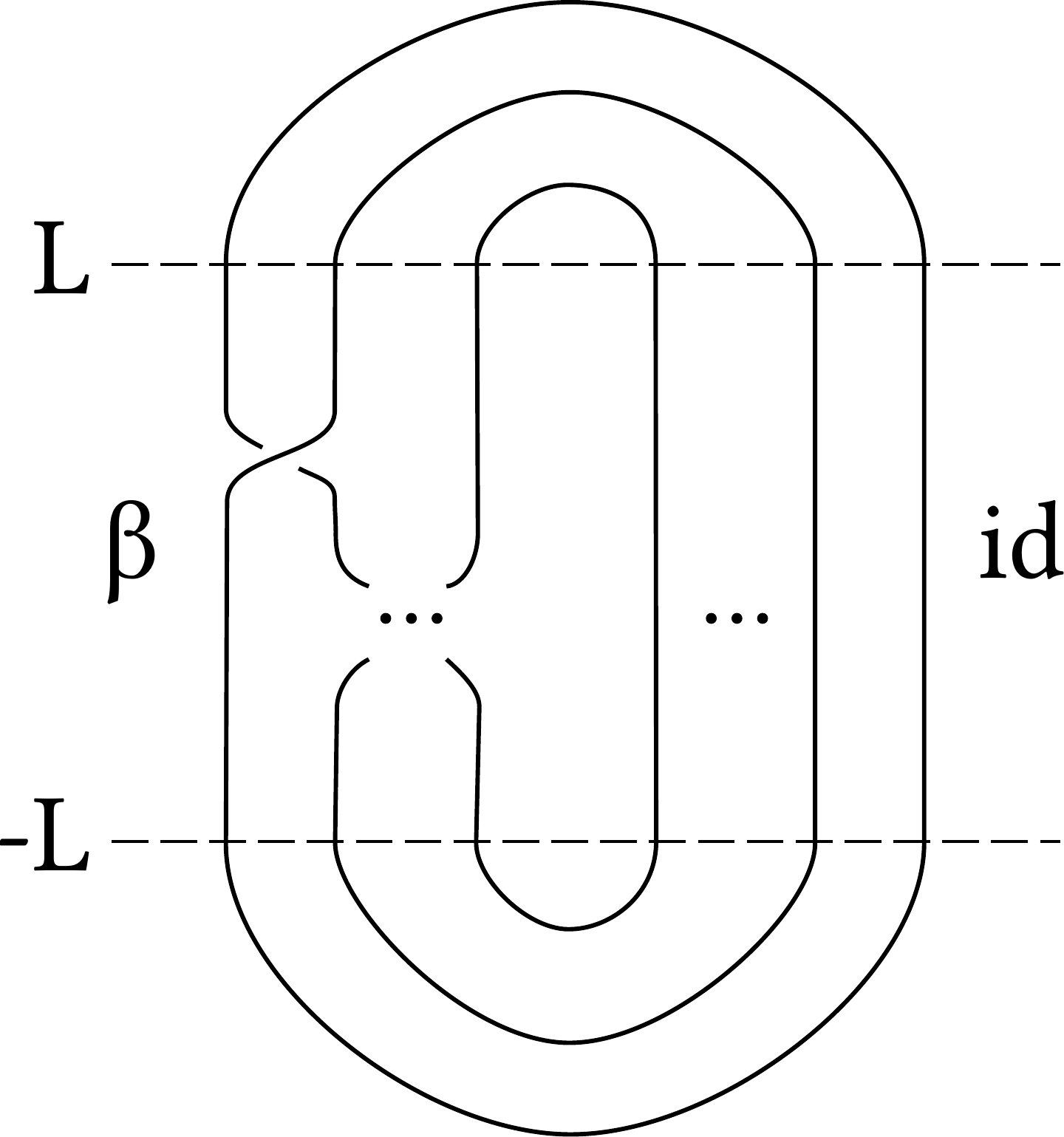}
\end{minipage}
\hspace{2cm}
\begin{minipage}{5cm}
\includegraphics[width=5cm, height=5cm, keepaspectratio]{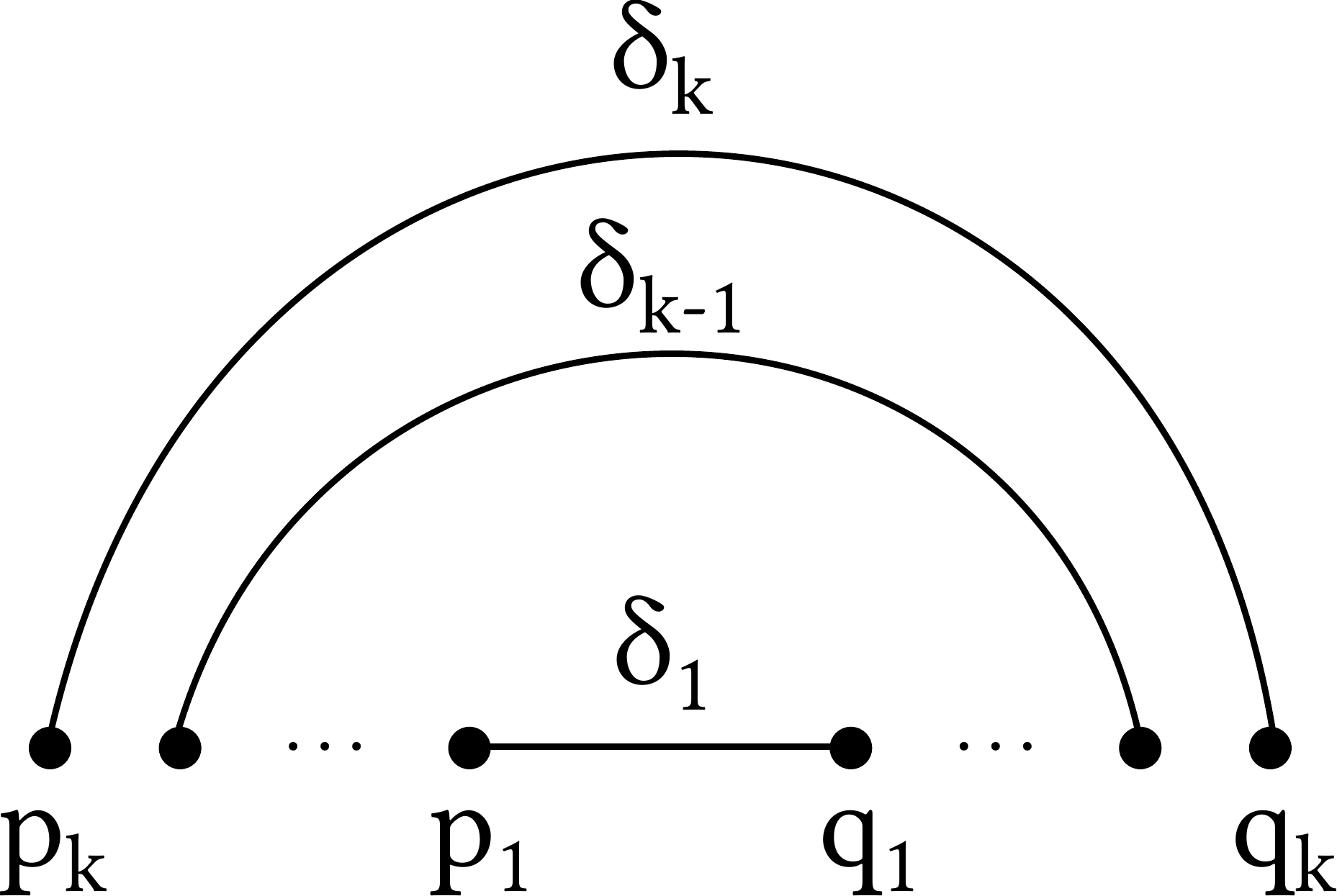}
\end{minipage}

\caption{Left: Every knot $K$ can be viewed as closure of a bipartite braid $\beta \times \id$ embedded in $[-L,L]_t \times \mathbb{C}$ by gluing in cups and caps. 
Right: Gluing instructions for cups and caps are captured by a crossingless matching $\mathfrak{m}$ consisting of $k$ disjoint arcs $\delta_i \subset \mathbb{C}$.}
\label{fig:khovanov-braid-closure}
\end{figure}

Let us denote the configuration space of $2k$ points that are partitioned into two individual sets of $k$ points by $\Conf_{k,k} \mathbb{C}$.
The closure of a braid $\beta$ is determined by the following data (cf. \autoref{fig:khovanov-braid-closure}).
First, an embedding of a bipartite braid of the form $\beta\times\id : [-L, L]_t \to \Conf_{k,k} \mathbb{C}$ into $[-L,L] \times \mathbb{C}$, where the identity braid on the second set of points is assumed to be constant.
And second, by a choice of crossingless matching between the two sets of points determined by $\beta\times \id$ at $t=-L$, and analogously for $t=L$.
To make this precise, assume the bipartite braid $\beta\times\id$ determines a collection of points $D=\{(p_1, \ldots, p_k), (q_1, \ldots q_k) \}$ at $t=L$.
A crossingless matching of $D$ is a collection of $k$ disjoint embedded arcs $\mathfrak{m} = (\delta_1, \ldots, \delta_k)$ in $\mathbb{C}$, with starting points $\delta_i(0)=p_i$ and endpoints $\delta_i(T) = q_i$.
The arcs $\delta_i$ determine which strands are glued together by cups below $t=-L$ (respectively, by caps above $t=L$) to make the open braid $\beta\times\id : [-L, L]_t \to \Conf_{k,k} \mathbb{C}$ into a closed knot in $\mathbb{R}_t \times \mathbb{C}$.

More abstractly, a crossingless matching determines an extension of $\beta\times\id$ into the singular locus of $\overline{\Conf_{k,k} \mathbb{C}} \simeq \overline{\Conf_{2k} \mathbb{C}}$.
Here we view the closure of the configuration space as a stratified space by identifying configurations of $k$ points with two identical points as a configuration of $k-1$ points, and so on.
The associated stratification is given by the inclusions $\Conf_{k-1} \mathbb{C} \subset \overline{\Conf_{k} \mathbb{C}}$, with lowest stratum $\Conf_{0} \mathbb{C} = \emptyset$.
An entrance path in a stratified space is a path that starts in a higher-dimensional stratum and can only transition into lower-dimensional strata or remain in the same stratum: it only ever enters lower-dimensional strata.

A single arc $\delta_i$ can be viewed as an entrance path $[0,1]\to \overline{\Conf_{k,k} \mathbb{C}}$ that starts in the top stratum $\Conf_{k,k} \simeq \Conf_{2k} \mathbb{C}$ and ends in the lower stratum $\Conf_{2k-1}\mathbb{C}$.
For this one identifies the arc with the path determined by $p_i(t) = \delta(t/2)$ and $q_i(t) = \delta_i (1-t/2)$.
Moreover, after two points have been matched, the two strands of the braid are closed off and for the purposes of describing the knot closure we can simply forget about the endpoint, further mapping $\Conf_{2k-1}\mathbb{C} \to \Conf_{2k-2}\mathbb{C}$ which is then identified with $\Conf_{k-1, k-1} \mathbb{C}$ in the obvious way.
By successively following the arcs in $\mathfrak{m}=(\delta_1, \ldots, \delta_k)$, a crossingless matching thus defines an entrance path that starts in $\Conf_{k,k}$ and ends in $\emptyset$.

\begin{example}
Consider the case of four points $\{ (p_1, p_2), (q_1,q_2) \}$ and an arc $\delta_1$ that connects $p_1$ and $q_1$ at $\delta_1(1)=r \in \mathbb{C}$.
Then this determines the following entrance path connecting $\Conf_{2,2}\mathbb{C}$ to $\Conf_{1,1}\mathbb{C}$:
\begin{align}
\scriptsize
\begin{matrix}
	\overline{\Conf_{2,2}\mathbb{C}} \simeq \overline{\Conf_{4}\mathbb{C}} &\supset& 
	\overline{\Conf_{3}\mathbb{C}} &\supset& 
	\overline{\Conf_{2}\mathbb{C}} \simeq \Conf_{1, 1} \mathbb{C}, \\[0.5em]
	\{(p_1(t),p_2), (q_1(t),q_2) \} &\to&
	\{ (r, p_2), (r,q_2) \} \simeq \{ r, p_2, q_2 \} &\to&
	\{p_2,q_2\} \simeq \{ (p_2), (q_2) \}.
\end{matrix}
\end{align}
\end{example}

Closing off an open braid by cups and caps imposes constraints on $k$-monopole configurations.
From the point of view of two individual monopoles, located at $p_i$ and $q_i$, the matching arc $\delta_i$ specifies that their two configurations must be identical at $\delta_i(1)$.
But this also means that they can't be too different for $\delta_i(1 - \epsilon)$.
In that way a crossingless matching $\mathfrak{m}$ forces monopole configurations near $t=\pm L$ to be elements of a compact Lagrangian subspace $\mathcal{L}_{\mathfrak{m}} \subset \mathcal{Y}_{\pi,\rho,D}$.
While we approach the problem from a slightly different perspective, the construction of $\mathcal{L}_{\mathfrak{m}}$ is equivalent to the one described by Seidel-Smith and Manolescu \cite{Seidel2004, Manolescu2007}.
We proceed by induction on the number of arcs $k$ in a given matching.

Start with $k=1$.
The matching $\mathfrak{m}$ contains a single arc $\delta$, matching $2$ points $\{(p),(q)\} \in \Conf_{1,1}\mathbb{C}$ that are labeled by the same partition $\pi_1 = \pi_2 = \pi$.
Interpret the arc as an entrance path $\delta:[0,1] \to \overline{\Conf_{1,1}\mathbb{C}}$, starting at $\delta(0) = (p,q) \in \Conf_{1,1}\mathbb{C}$ and ending at $\delta(1) = r \in \Conf_{1} \mathbb{C} $.
There is nothing special about the choice of $r$, so we pick $r=0$ for simplicity.
The Grothendieck-Springer fibration $\tilde\chi: \mathfrak{sl}(2N,\mathbb{C})^{[(N-1)^2 1^2]} \to \overline{\Conf_{1,1}\mathbb{C}}$ has singular fibers over configurations with $p=q$.
The singularity in the fibers corresponds to the loci of matrices with Jordan blocks larger than $1$.

\begin{figure}
\centering
\includegraphics[width=\textwidth, height=0.3\textheight, keepaspectratio]{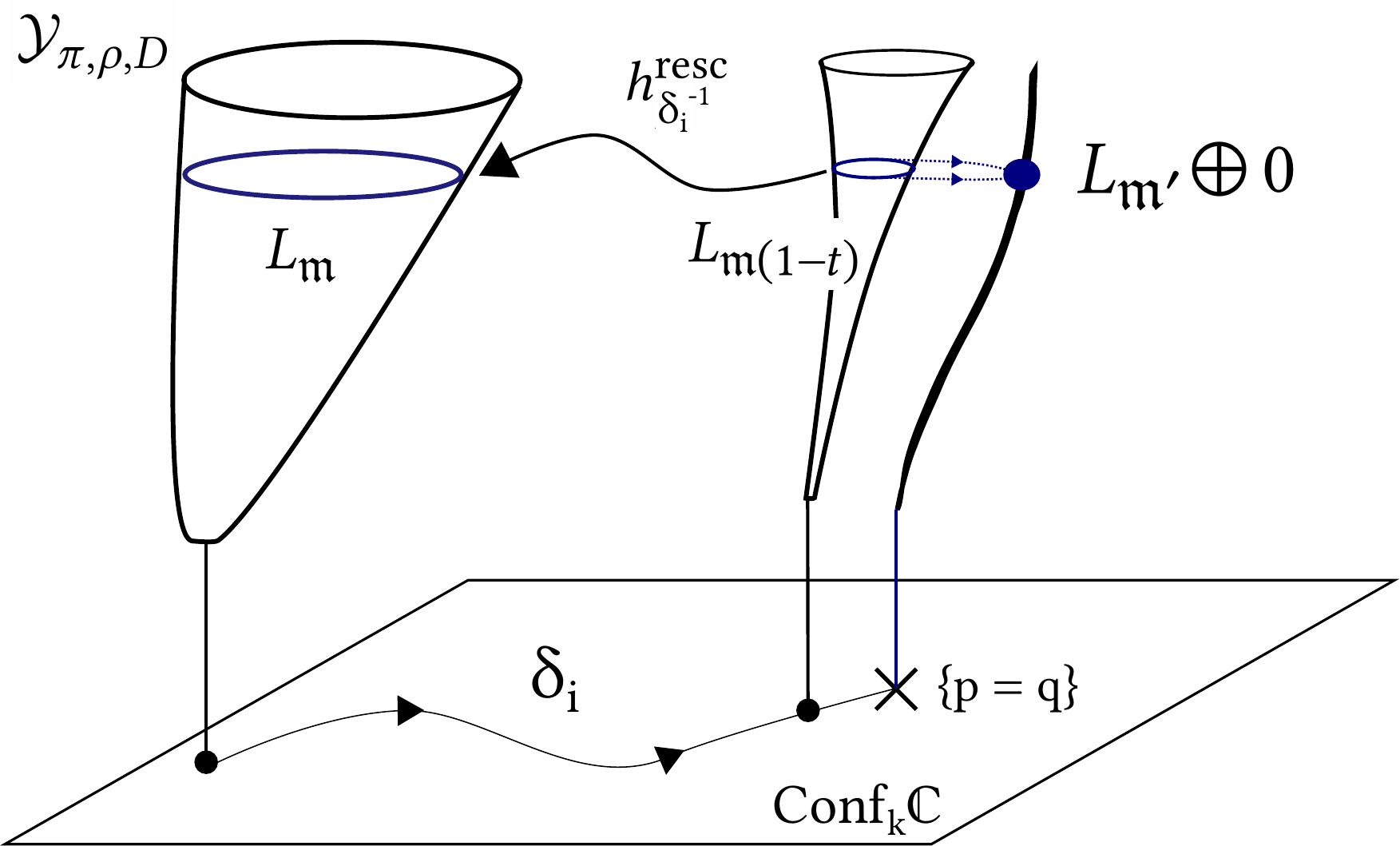}
\caption{Inductive construction of the vanishing cycle $\mathcal{L}_{\mathfrak{m}}$.
For a given arc $\delta_i$ of a crossingless matching $\mathfrak{m} = (\delta_1, \ldots \delta_k)$, }
\label{fig:khovanov-lagrangians}
\end{figure}

We use naive (as opposed to rescaled) parallel transport $h_\delta$ along $\delta$ to define, for sufficiently small $t\in[0,1]$, the following subsets of the fibers near the singular locus $p=q=0$:
\begin{align}
	\mathcal{L}_{1-t} := \set{ Y \in \mathcal{Y}_{[\pi^2], [N^2], \delta(1-t)} \suchthat
	\begin{matrix}
		\text{$h_{\delta\restr_{[0,s]}}$ is defined in a neighbourhood} \\
    \text{of $Y$ for all $s<1$ and} \\
		h_{\delta(s)}(Y) \stackrel{s\to 1}{\longrightarrow} D(r) \oplus D(r)\ = 0
	\end{matrix}
	}.
\end{align}
It follows from \cite[Sec. 4.3]{Manolescu2007} that $\mathcal{L}_t$ is diffeomorphic to the direct sum of two copies of $\mathbb{C} P^{N-1}$ when $\pi=[(N-1) 1]$ (which corresponds to a magnetic charge $\lambda = (1,0\ldots, 0)$ or equivalently a strand labeled by the fundamental representation).
Each of the two projective spaces arises as a quotient of an ordinary vanishing cycle $S^{2n-1}$ by an $S^1$ action.
For more general magnetic charges $\lambda$ it is expected that one obtains vanishing Grassmannians instead of vanishing projective spaces.

In any case, $\mathcal{L}_{1-t}$ is a Lagrangian submanifold of $\mathcal{Y}_{\pi, [N^2], \delta(1-t)}$.
We then use rescaled parallel transport ``backwards'' along $\delta$ to move $\mathcal{L}_{1-t}$ all the way to a subspace $\mathcal{L}_0$ in the initial fiber $\mathcal{Y}_{[\pi^2], [N^2], \delta(0)}$, as depicted in~\autoref{fig:khovanov-lagrangians}.

For the induction step, assume we are given the positions of $2k$ strands $D=\{(p_a), (q_a)\}_{a=1,\ldots, k} \in \Conf_{k,k} \mathbb{C}$, labeled by an admissible partition $\pi = [\pi_{p_1}\ldots\pi_{q_1}\ldots]$ of $2kN$, and a crossingless matching $\mathfrak{m} = (\delta_1, \ldots, \delta_k)$.
We are interested in the effect of matching points along $\delta_1:[0,1] \to \overline{\Conf_{k,k} \mathbb{C}}$, so we consider $\mathfrak{m}(t) = (\delta_1(t), \delta_2, \ldots \delta_k)$ with $m(0) = D$.
Denote by $D'$, $\pi'$, and $\mathfrak{m}'$ the objects that are obtained from $D$, $\pi$, and $\mathfrak{m}$ after removing $(p_1,q_1)$ along the arc $\delta_1$.

Observe that we can split
\begin{align}
	\mathfrak{sl}(2kN,\mathbb{C})^{[N^{2k} 1^{2k}]} = \mathfrak{sl}((2k-2)N,\mathbb{C})^{[N^{2k-2} 1^{2k-2}]} \oplus \mathfrak{sl}(N,\mathbb{C})^{[(N-1) 1]}.
\end{align}
The space $\mathcal{Y}_{\pi', [N^{2k-2}1^{2k-2}], D'}$ can be identified with a fiber of
\begin{align}
	\mathfrak{sl}((2k-2)N,\mathbb{C})^{[N^{2k-2} 1^{2k-2}]} \oplus \{0\} \stackrel{\tilde\chi}{\to} \Conf_{k-1,k-1}.
\end{align}
By assumption, we have a Lagrangian $\mathcal{L}_{\mathfrak{m}'} \subset \mathcal{Y}_{\pi', [N^{2k-2}1^{2k-2}], D'}$.
Using parallel transport, we can move this Lagrangian into the singular locus of $\mathcal{Y}_{\pi, [N^{2k}1^{2k}], \mathfrak{m}(1)}$.
By a relative version of the previous construction, one obtains for small $t$ a vanishing Lagrangian $\mathcal{L}_{\mathfrak{m}(1-t)} \subset \mathcal{Y}_{\pi, [N^{2k}], \mathfrak{m}(1-t)}$ that is diffeomorphic to $\mathcal{L}_{\mathfrak{m}'(1-t)} \oplus \mathcal{L}_{1-t}$.
Using rescaled parallel transport along $\mathfrak{m}(t)$, we move this to the fiber over $\mathfrak{m}(0)$, which yields the desired Lagrangian $\mathcal{L}_{\mathfrak{m}} \subset \mathcal{Y}_{\pi,\rho, D}$.

\begin{figure}
\centering
\includegraphics[width=0.8\textwidth,height=0.4\textheight, keepaspectratio]{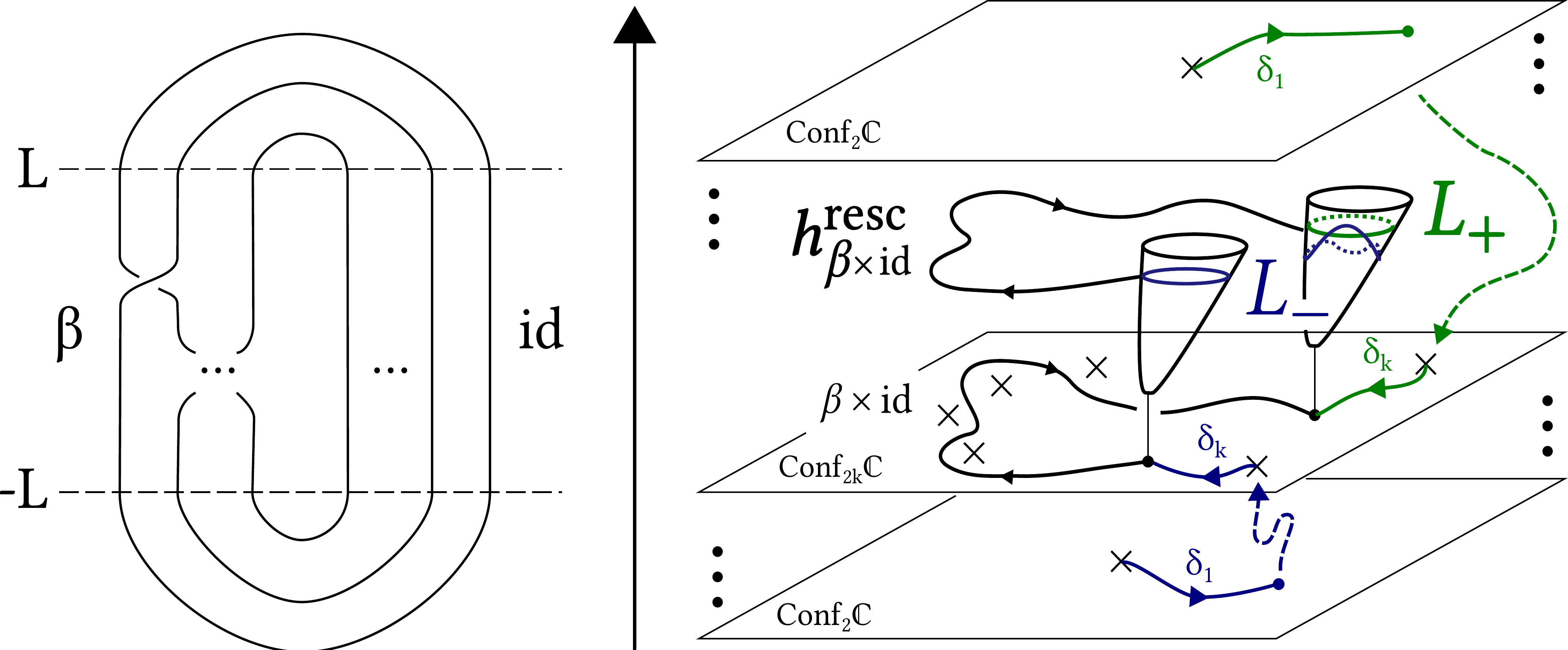}
\caption{Points in the intersection $h_{\beta\times\mathrm{id}}^{\mathrm{resc}} \mathcal{L}_- \cap \mathcal{L}_+$ determine horizontal lifts of the braid closure $\overline{\beta}$. Each horizontal lift determines a way to glue solutions of the EBE to a global monopole configuration along $K\subset \partial \mathbb{R}_t \times \mathbb{C}$.}
\label{fig:khovanov-symplectic-khovanov-schematics}
\end{figure}

Using this construction, we attach two Lagrangian vanishing spaces $\mathcal{L}_-$, $\mathcal{L}_+ \subset \mathcal{Y}_{\pi,\rho,D}$ to a choice of crossingless matchings at $t=-L$ and $t=L$, respectively.
Parallel transport of $\mathcal{L}_-$ along the braid provides a horizontal path in the Grothendieck-Springer fibration and any point in the intersection $h_{\beta\times\mathrm{id}}^{\mathrm{resc}} \mathcal{L}_- \cap \mathcal{L}_+$ corresponds to a path that connects the bottom part of the path to the top one, see~\autoref{fig:khovanov-symplectic-khovanov-schematics}.
The pre-image of such a path under the bundle map $\varUpsilon$ in~\eqref{eq:khovanov-effective-triple-to-monopole-moduli} then contains at least one non-vertical family of effective triples, which in turn determines a solution of the decoupled Kapustin-Witten equations.

\section{The Floer Differential and Symplectic Khovanov Homology}
\label{sec:khovanov-homology}

In this section we describe implications of our discussion for Witten's gauge theoretic approach to Khovanov homology and formulate a second main conjecture.
For this we return to the full decoupled Haydys-Witten equations on $M^5 = C \times \Sigma \times \mathbb{R}^+_y$, where $C = \mathbb{R}_s \times \mathbb{R}_t$ with holomorphic coordinate $w=s+it$, and $\Sigma = \mathbb{C}$ with holomorphic coordinate $z$.

Let us start with a brief reminder of Haydys-Witten Floer homology $HF_{\pi/2} (W^4)$ in the context of Khovanov homology~\cite{Witten2011, Bleher2024b}.
Assume $W^4 = \mathbb{R}_t \times \mathbb{C} \times \mathbb{R}^+_y$ and that we are given a knot $K \subset \partial W^4 = \mathbb{R}_t \times \mathbb{C}$.
The Floer cochain complex is defined to be the abelian group generated by solutions of the $\theta=\pi/2$-version of the Kapustin-Witten equations that satisfy Nahm pole boundary conditions with knot singularities along $K$:
\begin{align}
	CF_{\pi/2}([W^4;K]) := \bigoplus_{x \in \mathcal{M}^{\mathrm{KW}}(W^4)}  \mathbb{Z} [x].
\end{align}
Equip $M^5 = \mathbb{R}_s \times W^4$ with the non-vanishing vector field $v=\partial_y$ and extend $K$ to the $\mathbb{R}_s$-invariant surface $\Sigma_K = \mathbb{R}_s \times K$ in the boundary of $M^5$.
Then there is an associated Floer cohomology group $HF_{\pi/2}([W^4;K]) := H(CF_{\pi/2} , d_v )$ with respect to the Haydys-Witten Floer differential defined by
\begin{align}
	d_{v\/} x = \sum_{\mu(x,y)=1} m_{xy} \cdot y.
\end{align}
Here $m_{xy}$ is the signed count of Haydys-Witten instantons on $M^5$, subject to the following conditions
\begin{align}
	\renewcommand{\arraystretch}{1.3}
	\left\{
	\begin{array}{l}
		\HW[v](A,B) = 0, \\
		\lim_{s\to -\infty} (A,B) = x,\ \lim_{s\to+\infty} (A,B) = y, \\
		\text{$(A,B)$ satisfy regular Nahm pole boundary conditions}\\[-0.2em]
		\qquad \text{with knot singularities along $\Sigma_K$.}
	\end{array}
	\right.
\end{align}
Witten's gauge theoretic approach to Khovanov homology is the proposal that $HF_{\pi/2}([W^4, K], v=\partial_y)$ coincides with Khovanov homology.

The manifold $M^5 = C \times \Sigma \times \mathbb{R}^+_y$ with $C= \mathbb{C}_w$ and $\Sigma=\mathbb{C}_z$ is special in two important ways.
First, there are no non-trivial Vafa-Witten solutions at $y\to\infty$, such that there is an instanton grading $HF^\bullet_{\pi/2}([\mathbb{R}_t\times \mathbb{C} \times\mathbb{R}^+_y, K], v=\partial_y)$.
In fact, physical considerations suggest that there actually is a bigrading, where the second grading is related to an absolute version of the Maslov index \cite{Witten2011, Bleher2024b}.
Second, looking at conditions (A1)~-~(A4) of~\cite{Bleher2023b} and writing $B=B^{\mathrm{NP}}+b$, with component $b_z = b_2+ib_3$, we find that in the current situation any Haydys-Witten solution that satisfies $D_{\bar w\/} b_z = \mathcal{O}(y^2)$ is already a solution of the decoupled Haydys-Witten equations.
While it is unclear if there are solutions for which the order $y^1$ term of $b_z$ is not holomorphic, we take this observation as strong motivation to assume that all Haydys-Witten instantons are solutions of the decoupled equations.

Regarding the Floer chains, \autoref{sec:khovanov-braids-to-knots} suggests that solutions of the decoupled Kapustin-Witten equation on $W^4 = \mathbb{R}_t \times \mathbb{C} \times \mathbb{R}^+_y$ with knot singularities are given by intersections of certain Lagrangian subspaces of $\mathcal{Y}_{\pi,\rho,D}$.
To determine the Floer differential it remains to describe Haydys-Witten instantons on $M^5$ that interpolate between Kapustin-Witten solutions at $s\to\pm\infty$.
Observe, again, that the decoupled Haydys-Witten equations~\eqref{eq:khovanov-decoupled-haydys-witten-4D} contain the extended Bogomolny equations $[\mathcal{D}_i, \mathcal{D}_j] = 0$.
As before, the latter define a family of effective triples, now parametrized by $C=\mathbb{C}_w$ and given by a map
\begin{align}
	u: \mathbb{C}_w \to \mathcal{M}_{(\mathcal{E},\varphi,L)} \; ,\ w \mapsto (\mathcal{E}(w), \varphi(w), L(w) ).
\end{align}
The remaining $\mathcal{G}_{\mathbb{C}}$-invariant equations $[\mathcal{D}_0, \mathcal{D}_i] = 0$ become equivalent to Cauchy-Riemann equations $\nabla_{\bar w\/} u = 0$, so $u$ is a pseudo-holomorphic disc.

\begin{remark}
Notably, every ``large'' gauge transformation $g \in \mathcal{G}_\mathbb{C}(M^5)$, i.e. such that $g(w, z, y=0)$ is non-zero, that is holomorphic in $w$ gives rise to a holomorphic disc $u=g \cdot (\mathcal{E},\varphi, L)$.
Consequently, solutions of the EBE that become indistinguishable in homology whenever they are related by such a holomorphic ``large'' gauge transformation.
This observation may provide an a posteriori justification for passing from the infinite dimensional moduli space of effective triples $\widehat{\mathcal{M}}_{(\mathcal{E},\varphi,L)}$ to the finite dimensional model space $\mathcal{Y}_{\pi,\rho,D}$, as was mentioned previously.
Making this rigorous requires a detailed analysis of the boundary configurations of effective triples (or equivalently EBE-solutions) and their holomorphic gauge equivalence classes under ``large'' gauge transformations.
\end{remark}

We now need to embark on a short detour and briefly review the definition of Lagrangian intersection Floer theory \cite{Floer1988a} (omitting virtually all technical details).
Let $(M,\omega)$ be a K\"ahler manifold and consider closed connected Lagrangian submanifolds $L, L' \subset M$.
Let $O_x$ denote the orientation group of the point $x$, defined as the group that is generated by the two possible orientations of $x$ together with the relation that their sum is zero (thus, $O_x$ is non-canonically isomorphic to $\mathbb{Z}$).
The Lagrangian intersection cochain complex is defined to be the abelian group
\begin{align}
	CF(L,L') := \bigoplus_{x \in L\cap L'} O_x.
\end{align}
From this one defines Lagragian intersection Floer cohomology $HF(L, L') := H(CF (L, L'), d_J )$ as cohomology with respect to the differential
\begin{align}
	d_J x = \sum_{y \in L\cap L'} n_{xy} y.
\end{align}
Here $n_{xy}$ is a signed count of pseudoholomorphic discs, with respect to a generic smooth family of $\omega$-compatible almost complex structures $\{J_t\}$, that are subject to the following conditions.
\begin{align}
	\renewcommand{\arraystretch}{1.2}
	\left\{
	\begin{array}{l}
		u: \mathbb{R}_s \times [0,1]_t \to M, \\
		\partial_s u + J_t(u) \partial_t u = 0, \\
		u(s,0) \in L,\ 	u(s,1) \in L', \\
		\lim_{s\to-\infty} u(s,\cdot) = x,\ \lim_{s\to\infty} u(s,\cdot) = y.
	\end{array}
	\right.
\end{align}

An example of a Lagrangian intersection Floer theory that is of particular relevance to us is symplectic Khovanov-Rozansky homology of a knot $K$, developed by Seidel-Smith and Manolescu \cite{Seidel2004, Manolescu2007}.
The starting point are Lagrangian submanifolds $\tilde{\mathcal{L}}_\pm$ of $\mathfrak{sl}(kN,\mathbb{C})$ that are constructed in exactly the same way as the vanishing spaces $\mathcal{L}_\pm$ above.
However, there is a crucial difference between the fibration that is used here and the one defined more carefully by Seidel-Smith and Manolescu.
From the perspective of physics it was natural to encode each of the $2k$ monopoles of the braid closure $\bar{\beta} = K$ in an individual $\mathfrak{sl}(N,\mathbb{C})$-block, leading to a construction of $\mathcal{L}_\pm$ in $\mathfrak{sl}(2kN, \mathbb{C})$.
In contrast, symplectic Khovanov-Rozansky homology is more efficient and utilizes the thin eigenvalues to encode the position of the $S^1$-invariant ``returning'' strands of the braid closure $\bar{\beta}$, such that the associated Lagrangian submanifolds $\tilde{\mathcal{L}}_\pm$ live in $\mathfrak{sl}(kN,\mathbb{C})$.

Given $\tilde{\mathcal{L}}_\pm$, symplectic Khovanov-Rozansky for a braid $\beta$ on $k$ strands is defined as the Lagrangian intersection Floer theory of $h^{\mathrm{resc}}_{\beta\times\mathrm{id}} \tilde{\mathcal{L}}_-$ and $\tilde{\mathcal{L}}_+ \subset \mathfrak{sl}(kN)$.
\begin{definition}[Symplectic Khovanov Homology \cite{Seidel2004, Manolescu2007}]
\begin{align}
	\mathcal{H}^\bullet_{\mathrm{symp. Kh}}(K) := HF^\bullet ( h_{\beta\times\id}^{\mathrm{resc}} \tilde{\mathcal{L}}_-, \tilde{\mathcal{L}}_+).
\end{align}
\end{definition}
Seidel-Smith proved that symplectic Khovanov homology is a knot invariant for $\mathfrak{sl}(2, \mathbb{C})$ and Manolescu generalized the construction and proof to $\mathfrak{sl}(N,\mathbb{C})$.
It is expected that symplectic Khovanov homology coincides with a grading-reduced version of Khovanov-Rozansky homology.
In fact, this has been proven for $\mathfrak{sl}(2,\mathbb{C})$ by Abouzaid and Smith:
\begin{theorem}[\cite{Abouzaid2019}]
\label{thm:khovanov-symp-Kh-is-Kh}
Let $\mathfrak{g} = \mathfrak{sl}(2,\mathbb{C})$. 
Then for any oriented link $K$ one has an absolute grading and an isomorphism
\begin{align}
	\mathcal{H}_{\mathrm{symp. Kh}}^k(K) \simeq \bigoplus_{i-j = k} \mathcal{H}^{i,j}_{\mathrm{Kh}}(K).
\end{align}
\end{theorem}

Let us now return to Haydys-Witten instanton Floer theory.
Since solutions of the decoupled Kapustin-Witten equations are related to Lagrangian intersections (\autoref{sec:khovanov-braids-to-knots}), we conjecture that the Hay\-dys-\-Wit\-ten Floer complex can be replaced by
\begin{align}
	CF_{\pi/2}([\mathbb{C} \times \mathbb{R}^+_y; K]) = CF( \mathcal{L}_-, \mathcal{L}_+).
\end{align}
Moreover, by our explanations above, the Haydys-Witten Floer differential is determined by pseudo-holomorphic discs $u$ of width $2L\in\mathbb{R}$ that satisfy the following conditions.
\begin{align}
	\renewcommand{\arraystretch}{1.2}
	\left\{
	\begin{array}{l}
		u: \mathbb{R}_s \times [-L, L]_t \to \mathfrak{sl}^{[(N-1)^{2k}\ 1^{2k}]}, \\
		\partial_s u + i \partial_t u = 0, \\
		\lim_{s\to-\infty} u(s,\cdot) = x,\ \lim_{s\to\infty} u(s,\cdot) = y, \\
		u(s,-L) \in \mathcal{L}_-,\ 	u(s,+L) \in \mathcal{L}_+ .
	\end{array}
	\right.
\end{align}
The last condition is satisfied by virtue of $\mathbb{R}_s$-invariance of $\Sigma_K$.
In conclusion, we claim that Haydys-Witten Floer homology coincides with Lagrangian intersection homology
\begin{align}
	HF_{\pi/2}\big( [\mathbb{R}_t \times \mathbb{C} \times \mathbb{R}^+_y; K] \big) = HF\big( h_{\beta\times\id}^{\mathrm{resc}} \mathcal{L}_- , \mathcal{L}_+ \big).
\end{align}
While the right hand side is not symplectic Khovanov-Rozansky homology, the difference is only in handling the constant parts of the knot closure $\bar{\beta} = \beta \times \id$.
Motivated by He and Mazzeo's classification of $S^1$-invariant Kapustin-Witten solutions, we propose that the effect of the $k$ $S^1$-invariant monopoles supported on $\id$ can be neglected.
Using this, we arrive at the second main conjecture of this thesis.
\begin{conjecture}
\label{conj:khovanov-HF-is-symp-Kh}
Haydys-Witten Floer homology of $[\mathbb{R}_t \times \mathbb{C} \times \mathbb{R}^+_y; K]$ coincides with symplectic Khovanov-Rozansky homology
\begin{align}
	HF^\bullet_{\pi/2}\big( [\mathbb{R}_t \times \mathbb{C} \times \mathbb{R}^+_y; K] \big) = \mathcal{H}^\bullet_{\text{symp. Kh}}(K).
\end{align}
\end{conjecture}
As a consequence of \autoref{thm:khovanov-symp-Kh-is-Kh}, the ideas leading up to \autoref{conj:khovanov-HF-is-symp-Kh} provide a novel approach to proofing a slightly weakend version of Witten's original gauge-theoretic construction of Khovanov homology:
\begin{theorem}
If \autoref{conj:khovanov-HF-is-symp-Kh} is true, then Haydys-Witten Floer theory with $G=SU(2)$ coincides with a grading reduced version of Khovanov homology.
\begin{align}
	HF^k_{\pi/2}\big( [\mathbb{R}_t \times \mathbb{C} \times \mathbb{R}^+_y; K] \big) \simeq \bigoplus_{i-j = k} \mathrm{Kh}^{i,j}(K).
\end{align}
\end{theorem}

\printbibliography[heading=bibintoc]

\end{document}